\documentclass[12pt,english]{article}

\usepackage[latin9]{inputenc}
\usepackage{geometry}
\usepackage[greek,USenglish]{babel}
\usepackage{verbatim}
\usepackage{prettyref}
\usepackage{amsmath}
\usepackage{amssymb}
\usepackage{subfig}
\usepackage{graphicx}
\usepackage{setspace}
\usepackage{esint}
\usepackage{marvosym}
\usepackage{tikz}
\usepackage{color}
\definecolor{darkgreen}{rgb}{0,0.5,0}
\definecolor{darkblue}{rgb}{0,0,0.6}
\definecolor{purple}{rgb}{0.4,.2,0.7}
\usepackage[colorlinks=true,citecolor=darkgreen,linkcolor=darkblue,urlcolor=purple]{hyperref}

\numberwithin{equation}{section}
\numberwithin{figure}{section}
\numberwithin{table}{section}

\begin{document}

\vspace*{1.5cm}
\begin{center}
{ \LARGE {\textsc{Supergoop Dynamics}}} 
\vspace*{1.5cm}
 
Dionysios Anninos$^{*}$, Tarek Anous$^{\dag}$,  Frederik Denef$^{{\dag},\text{\Cross}}$, George Konstantinidis$^{*}$ and \\ Edgar Shaghoulian$^{*}$
\\
\vspace*{1cm}
{\it $^*$ Department of Physics, Stanford University}\\
{\it $^{\dag}$ Center for the Fundamental Laws of Nature, Harvard University}\\
{\it $^\text{\Cross}$ Institute for Theoretical Physics, University of Leuven}
\vspace*{0.5cm}
\end{center}


\begin{abstract}

We initiate a systematic study of the dynamics of multi-particle systems with supersymmetric Van der Waals and electron-monopole type interactions. The static interaction allows a complex continuum of ground state configurations, while the Lorentz interaction tends to counteract this configurational fluidity by magnetic trapping, thus producing an exotic low temperature phase of matter aptly named supergoop. Such systems arise naturally in $\mathcal{N}=2$ gauge theories as monopole-dyon mixtures, and in string theory as collections of particles or black holes obtained by wrapping D-branes on internal space cycles. After discussing the general system and its relation to quiver quantum mechanics, we focus on the case of three particles. We give an exhaustive enumeration of the classical and quantum ground states of a probe in an arbitrary background with two fixed centers. We uncover a hidden conserved charge and show that the dynamics of the probe is classically integrable. In contrast, the  dynamics of one heavy and two light particles moving on a line shows a nontrivial transition to chaos, 
which we exhibit by studying the Poincar\'e sections. Finally we explore the complex dynamics of a probe particle in a background with a large number of centers, observing hints of ergodicity breaking. We conclude by discussing possible implications in a holographic context.  
%
%
%

\end{abstract}

\newpage

\tableofcontents


\section{Introduction}

\subsection{String glasses}

In the macroscopic world, where everything flows and nothing abides, mayhem and disorder rule. Most dynamical systems are not integrable but chaotic. Most materials are not crystalline but amorphous. They are glasses, eternally in search of their true equilibrium. Although glassy systems are still notoriously resistant to fundamental theoretical understanding, much progress has been made over the past thirty years \cite{leuzzi}, and remarkable organizing principles have been uncovered. 
A beautiful early example is given by the simplest toy model of a spin glass, the Sherrington-Kirkpatrick model, for which the space of low temperature configurations organizes itself in a hierarchic treelike fashion  \cite{spinglassbook,Denef:2011ee}. 

String theory presents us with similarly, if not more, complex systems, on microscopic scales. These manifest themselves when studying the vast microstates of black holes or the vast number of flux compactifications, to give but two examples. Consider for instance (as reviewed in detail in \cite{Denef:2011ee}), a ``3-charge'' D4-brane wrapped on a smooth four-cycle $\Sigma$ inside a six-torus $T^6 = (T^2)_1 \times (T^2)_2 \times (T^2)_3$, and bound to $n$ pointlike D0-branes. Denote the number of intersection points of $\Sigma$ with the sub-tori $(T^2)_A$ by $P^A$, $A=1,2,3$ --- these are the D4-charges of the system. Then $\Sigma$ has triple self-intersection product $\mathcal{P}^3 \equiv 6 \, P^1 P^2 P^3$, and of order $\mathcal{P}^3$ worldvolume deformation and flux degrees of freedom. 

In the regime $n \gg \mathcal{P}^3$, the pointlike D0-brane degrees of freedom dominate the degeneracy $\Omega$ of supersymmetric ground states, and one easily computes:
\begin{equation}
S_{\rm micro} = \log \Omega  \approx 2\pi\sqrt{\frac{n \mathcal{P}^3}{6}}.
\end{equation}
This agrees with the Bekenstein-Hawking entropy of the D4-D0 black hole. In its 5-dimensional uplifted version \cite{Strominger:1996sh}, the computation is in essence an application of the Cardy formula in the context of the AdS$_3$/CFT$_2$ correspondence \cite{Strominger:1997eq}. 

However, away from the Cardy regime, i.e.\ when $\mathcal{P}^3 \gtrsim n$, the order $\mathcal{P}^3$ D4-degrees of freedom (deformations of $\Sigma$ and turning on fluxes) become entropically dominant and the counting problem becomes far more intricate: Each choice of worldvolume fluxes induces a different, highly complex potential on the moduli space of deformations of $\Sigma$, and the ground states of the system correspond to the minima of this vast D-brane landscape. On the other hand, on the black hole side, an exponentially large number of molecule-like, multicentered stationary black hole bound state configurations  appears \cite{Denef:2000nb}, all with the same total charge, and they entropically dominate the single centered black hole \cite{Denef:2007vg}. In \cite{Anninos:2011vn}, it was argued in a probe analysis that this picture persists at least metastably for nonextremal black holes, up to a critical temperature where the single-centered D4-D0 black hole regains dominance (see also \cite{Chowdhury:2011qu} for a closely related analysis). 

This transition between a single black hole and a zoo of metastable multi-black hole configurations is reminiscent of a structural glass transition and the rugged free energy landscape picture associated to the glass phase. Indeed, once in a particular multicentered configuration, it may take exponentially long for the system to find the most entropic ``true equilibrium'' configuration, as it proceeds through exponentially suppressed thermal and quantum tunneling, and furthermore far from the true equilibrium state, the preferred direction of local relaxation processes is likely to push the system into the direction other highly stable quasi-equilibrium states instead of the true maximal entropy equilibrium configuration.

\subsection{Supergoop}

When sufficiently far separated and moving slowly close to a ground state configuration, the black hole constituents can be thought of as pointlike particles, moving in an approximately flat background, interacting with each other through specific static and velocity dependent interactions. These effective inter-particle interactions are highly constrained by the fact that these BPS systems preserve four supercharges: A nonrenormalization theorem \cite{Denef:2002ru} implies that once a metric has been fixed on the configuration space, the static and first order velocity dependent interactions are of a fixed form. For the flat metric: 
\begin{equation} \label{gooppotential}
 H = \sum_{p=1}^N \frac{1}{2 m_p} \left[ \left( {\bf p}_p - {\bf A}_p \right)^2 + \left( \sum_{q=1}^N \frac{\kappa_{pq}}{|{\bf x}_p-{\bf x}_q|} + \theta_p \right)^2 \right]  + \mbox{fermions} \, ,
\end{equation}
where ${\bf A}_p$ is the vector potential produced at ${\bf x}_p$ by a collection of Dirac monopoles of charge $\kappa_{pq}$ at positions ${\bf x}_q$.  The coefficients $\kappa_{pq} = - \kappa_{qp}$ equal the electric-magnetic symplectic products between the charges of the centers $p$ and $q$. The parameters $\theta_p$ and masses $m_p$ are fixed by the BPS central charge of each center. If the configuration space metric is not flat, the $m_p$ may depend on ${\bf x}_p$.

As a result of this nonrenormalization theorem, exactly the same supersymmetric Lagrangians also appear in very different physical contexts where four supercharges are preserved and the low energy degrees of freedom can be identified with spatial positions. One example is a mixture of well-separated elementary particles obtained by wrapping D-branes around various internal cycles of a Calabi-Yau manifold, interacting with each other through gravitational, vector and scalar interactions. Clearly this can be viewed as an extreme limit of the multi-black hole systems considered above, where the dyonic black holes have been replaced by dyonic particles. Another example are monopoles and dyons in ${\mathcal N}=2$ Yang-Mills theories \cite{Kim:2011sc}. 

A more remote example \cite{Denef:2002ru} is a collection of space-localized wrapped D-branes at weak string coupling in the \emph{substringy} distance regime. Their low energy degrees of freedom are given by a $(0+1)$-dimensional supersymmetric quiver quantum mechanics, with a position 3-vector and a $U(1)$ gauge symmetry for each singly wrapped brane (identified with the nodes of the quiver) and the lightest brane-brane stretched open string modes represented as bifundamental oscillator degrees of freedom (identified with the arrows of the quiver). When the branes are all well separated, i.e.\ when the quiver theory is on the Coulomb branch, the open string modes become very massive and can be integrated out. Again, the resulting effective theory for the position degrees of freedom must necessarily be of the form (\ref{gooppotential}) fixed by supersymmetry. The coefficients $\kappa_{pq}$ are now identified with the net number of arrows between two nodes. 

Thus this type of supersymmetric multi-particle mechanics appears in many contexts, in widely different regimes. 
Much effort has been put into understanding and counting the supersymmetric ground states of such systems, in part because of their key role in physics derivations of BPS wall-crossing formulae \cite{Denef:2002ru,Denef:2007vg,Dimofte:2009bv,Andriyash:2010qv,Manschot:2010qz}. 
However, little has been said about excitations or dynamics for these systems. There are a few exceptions: \cite{Avery:2007xf} studied the classical and quantum dynamics of the two-particle system and found it was integrable, and in \cite{Anninos:2011vn} the persistence of the black hole molecular configurations at finite temperature was studied. However, no studies of multi-particle dynamics or statistical mechanics have been done so far. In this paper we wish to take steps in these directions. 

Besides the motivation for understanding D-brane and black hole statistical mechanics and their potentially interesting interpretation as holographic glasses, such studies would also be of intrinsic interest, as these systems are rather unusual in several aspects. Due to the special form of the potential (\ref{gooppotential}), an $N$-particle bound state will have a $2(N-1)$-dimensional moduli space of zero energy ground state configurations folded in a very complicated way into the $3(N-1)$-dimensional full configuration space (factoring out the center of mass).  Naively one might therefore think that even at very low temperatures, the system would behave like a liquid, exploring large parts of the configuration space by flowing along the continuous valley of minimal energy configurations. As a simple example consider the case $N=2$. The particle distance is fixed and the moduli space is a sphere. One might think a probability density initially localized near a point on that sphere would quickly diffuse out over it. However, due to the effective electron-monopole Lorentz interaction between the particles, this is not quite right, as diffusion is obstructed by magnetic trapping. Another way of understanding this is conservation of angular momentum: Monopole-electron pairs carry intrinsic spin directed along their connecting axis, of magnitude equal to half their symplectic product. Hence they behave like gyroscopes. They resist changing direction; kicks will just cause them to wobble. 

Thus it is natural to hypothesize that these supersymmetric multi-particle systems behave partly like a liquid and partly like a solid at very low temperatures --- like goop. We will therefore refer to this peculiar state of matter as \emph{supergoop}.

\subsection{Dynamics}

Many times we study Hamiltonian systems that are classically integrable. For this to be the case one requires the existence of at least $N$ conserved charges (with all mutual Poisson brackets vanishing) for a system with a $2N$-dimensional phase space. Examples include single one-dimensional particles with an arbitrary potential, since the energy is conserved, and two body problems with a central force. Phase space trajectories of a classically integrable system will map $N$-dimensional tori. Generally, however, our system will not be classically integrable and we must confront a chaotic system. The simplest example of a chaotic system is the double pendulum which has a four-dimensional phase space with a single conserved quantity: the energy. One then studies the phase space trajectories of the double pendulum as a function of increasing energy. For sufficiently low energies the trajectories are constrained to live on a two-dimensional torus displaying quasi-integrable behavior. As the energy is increased this torus is deformed and eventually breaks apart into smaller tori. This process is seen to continue until there is no visible structure in the phase diagram, i.e. the system tends toward ergodicity. What is perhaps most remarkable about the transition to chaos is that it occurs in a gradual fashion in which smaller and smaller islands of regular behavior are spawned before the system loses all manifest structure.  A crucial question, especially for a system with many degrees of freedom, is quantifying when all ordered behavior disappears and how it depends on the parameters of the system (see for example \cite{hand,arnold}).

It is our aim to begin a systematic study of the dynamical aspects of the underlying brane system on the Coulomb branch. In this paper we mainly address the case of a three particle configuration. This is already a difficult non-integrable three-body problem. To render the problem tractable, we study first the classical ground states and subsequently the motion of a probe particle in a fixed background consisting of a two-centered bound state. Remarkably, we discover that the motion of the probe is classically integrable! This is due to the presence of an additional hidden conserved quantity and is somewhat reminiscent of the integrability of a Newtonian probe particle interacting gravitationally with a background of two fixed masses, as discovered by Euler and Jacobi. We then study the transition to chaos for a system of two probes in the presence of a heavy fixed particle, with the motion restricted to live on a line. This setup is directly analogous to the double pendulum, allowing us to exploit many of the tools developed for the study of the double pendulum. As for the double pendulum, by studying Poincar\'e sections we observe the formation of islands in phase space and the eventual transition to global chaos with no apparent structure in phase space. Finally, we begin to address the far more intricate dynamics of a system with a large number of centers. We provide a brief exposition of the behavior of a probe particle inside a molecule with a given number of fixed centers. We observe highly complex trajectories that become trapped for long times in the sense that they do not explore the entire molecule. \\
\\
\noindent {\bf Note added:} After submission, we became aware of \cite{Nersessian:2007gc, Krivonos:2006qd, Bellucci:2008rp, Bellucci:2007kx}, which elegantly prove the classical integrability that we discuss in Section \ref{integrable}.



\section{General Framework}

Consider a system of branes wrapped on the cycles of a six-dimensional compact space, such that they are pointlike in the non-compact $(3+1)$-dimensions. The interactions between them are governed by strings whose ends reside on the branes themselves. The low energy physics is governed by an $\mathcal{N} = 4$ supersymmetric quiver quantum mechanics \cite{Douglas:1996sw}. The nodes of the quivers have gauge groups associated to them and the low energy string degrees of freedom are chiral multiplets transforming in the bifundamental between two given nodes. The position degrees of freedom $\textbf{x}_p$ of the branes are scalars in the vector multiplets of the gauge groups. It was shown in \cite{Denef:2002ru} that if we study the Coulomb branch of the branes, i.e. integrate out the massive chiral multiplets, a non-trivial potential is generated which governs the dynamics of the $\textbf{x}_p$. The system of branes allows for a large family of bound states with fixed equilibrium distances as classical ground states. If the branes move too close the stretched strings become light and even tachyonic and the system enters the Higgs phase.\footnote{In fact, the weak string coupling limit $g_s \to 0$ always pushes the system to the Higgs phase.} On the other hand, if there is a sufficiently large number $n_B$ of branes placed at a single point, such that the product of the string coupling $g_s$ and $n_B$ becomes large, the system is best described by closed string exchange and hence supergravity. In what follows we will consider the theory of supersymmetric multiparticle mechanics describing the Coulomb branch of the brane system. This has the advantage of being a simple setup which is interesting in and of its own right while reproducing many of the features of the multicentered configurations that exist in supergravity. 

\subsection{Supersymmetric multiparticles}

The theory we consider is the multiparticle supersymmetric mechanics, which we refer to as {\it supergoop}, studied for example in \cite{Denef:2002ru,Denef:2000nb,Avery:2007xf,Diaconescu:1997ut,D'Hoker:1985et,D'Hoker:1986uh,D'Hoker:1985kb,Feher:1989xw,Feher:1988th,Bloore:1992fv,Coles:1990hr,Horvathy:2006hx,Ivanov:2002pc}.  We simply state the Lagrangian of the system, referring to \cite{Denef:2002ru} for details.

The supergoop Lagrangian is given by:
\begin{equation}\label{multipLag}
L=\sum_p \frac{m_p}{2}\left(\dot{\mathbf{x}}_p^2+D_p^2+2i\bar{\lambda}_p\dot{\lambda}_p\right)+
\sum_p\left(-U_pD_p+\mathbf{A}_p \cdot \dot{\mathbf{x}}_p\right)+\sum_{p,q}\left(C_{pq}\bar{\lambda}_p\lambda_q+\mathbf{C}_{pq}\cdot\bar{\lambda}_p\boldsymbol\sigma\lambda_q\right)~,
\end{equation}
where $\lambda_p$ is the fermionic superpartner to $\mathbf{x}_p$. The $D_p$ fields are auxiliary non-dynamical scalars. 
We have defined the functions:
\begin{equation}
U_p = \sum_q\frac{\kappa_{pq}}{2r_{pq}}+\theta_p~, \quad
\mathbf{A}_p = -\frac{1}{2}\sum_q\kappa_{pq}\left[\mathbf{A}^d(\mathbf{r}_{pq})+\mathbf{A}^d(\mathbf{r}_{qp})\right]\label{vecPot}~,
\end{equation}
where:
\begin{equation}
\mathbf{A}^d(\mathbf{x})=\frac{- y}{2r(z\pm r)}\hat{x}+\frac{ x}{2r(z\pm r)}\hat{y}
\end{equation}
is the vector potential of a single magnetic monopole of unit charge at the origin and $\mathbf{r}_{pq}\equiv\mathbf{x}_p-\mathbf{x}_q$. For the Lagrangian to be supersymmetric we further require $\kappa_{pq}=-\kappa_{qp}$. In the quiver quantum mechanics context, the $\kappa_{pq}$ are the number of bifundamentals connecting two nodes. The supercharges are given by:
\begin{equation}\label{eq:Supercharges}
Q_\alpha = -\sum_p i\,{\boldsymbol{\sigma}}^{\gamma}_\alpha \lambda^p_\gamma \cdot \left( \mathbf{p}_p - \textbf{A}_p \right) + \lambda^p_\alpha U_p~, \quad
\bar{Q}^\beta = \sum_q i\,{\boldsymbol{\sigma}}^{\beta}_\gamma \bar{\lambda}^{\gamma}_{q} \cdot \left( \mathbf{p}_q - \textbf{A}_q \right) - \bar{\lambda}^{\beta}_{q} U_q~.
\end{equation}
The Weyl spinors obey $(\lambda_\alpha)^* \equiv \bar{\lambda}^{\alpha}$ and the $\boldsymbol{\sigma}^\beta_\alpha$ are the usual Pauli matrices. The Hamiltonian of our system is defined as:
\begin{equation}\label{hamiltonian}
H=\sum_p \mathbf{p}_p\cdot \dot{\mathbf{x}}_p-L~, \quad \mathbf{p}_p \equiv m_p\dot{\mathbf{x}}_p+\mathbf{A}_p~.
\end{equation} 
Upon integrating out the the $D$-terms we find:
\begin{equation}\label{ham2}
H = \frac{1}{2 m_p} \sum_p \left[ (\mathbf{p}_p-\mathbf{A}_p)^2 +{U_p^2} \right]+ \sum_{p<q}\frac{\kappa_{pq}}{2 r_{pq}^3} \textbf{r}_{pq} \cdot \bar{\lambda}_{pq} {\boldsymbol\sigma} \lambda_{pq}~,
\end{equation}
where $\lambda_{pq} \equiv \lambda_p - \lambda_q$. Notice that the particle interactions include velocity dependent forces. Furthermore, the system has three-body interactions due to the appearance of $U_p^2$ in $H$.

As usual our Hamiltonian $H$ is related to the supercharges as $\{ Q_\alpha,\bar{Q}^\beta \}_{D.B.}  = -2i \delta_\alpha^\beta H$. However this is most easily checked in the quantum mechanics context where $\mathbf{p}_p\rightarrow-i\nabla_p$ and we replace the above Dirac bracket relation with the anticommutation relation:
\begin{equation}
\{ Q_\alpha,\bar{Q}^\beta \}  = 2 \delta_\alpha^\beta H~,
\end{equation}
where $\{ \lambda^p_\alpha, \bar{\lambda}^{\beta}_{q}\} = m^{-1}_p \delta^{p}_{q} \delta^\alpha_\beta$. 

\subsection{Classical features and multicentered black holes}

To study the classical properties of this theory, we can turn the fermionic fields off and only consider the bosonic part of~(\ref{multipLag}). 
Static BPS configurations occur when $U_p=0$ for all $p$, i.e. when:
\begin{equation}\label{MultipBoundState}
\sum_q \frac{\kappa_{pq}}{2r_{pq}}=-\theta_p~, \quad \forall \; \; p~.
\end{equation}
Taking the sum over $p$ of~(\ref{MultipBoundState}) we find that the $\theta_p$'s must satisfy: $\sum_p\theta_p=0$. 
\begin{figure}
\begin{center}
\includegraphics[height=80mm]{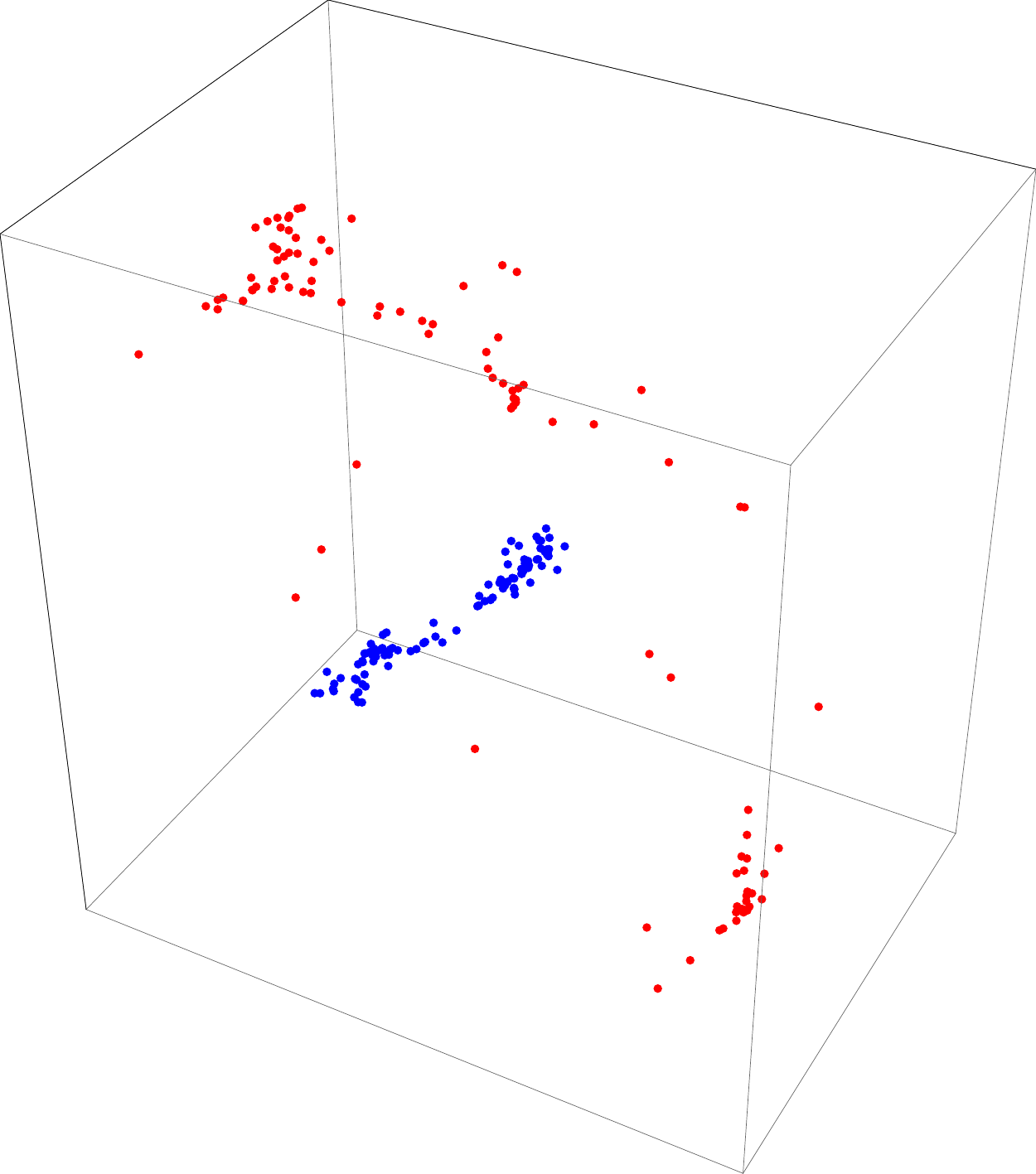}
\includegraphics[height=80mm]{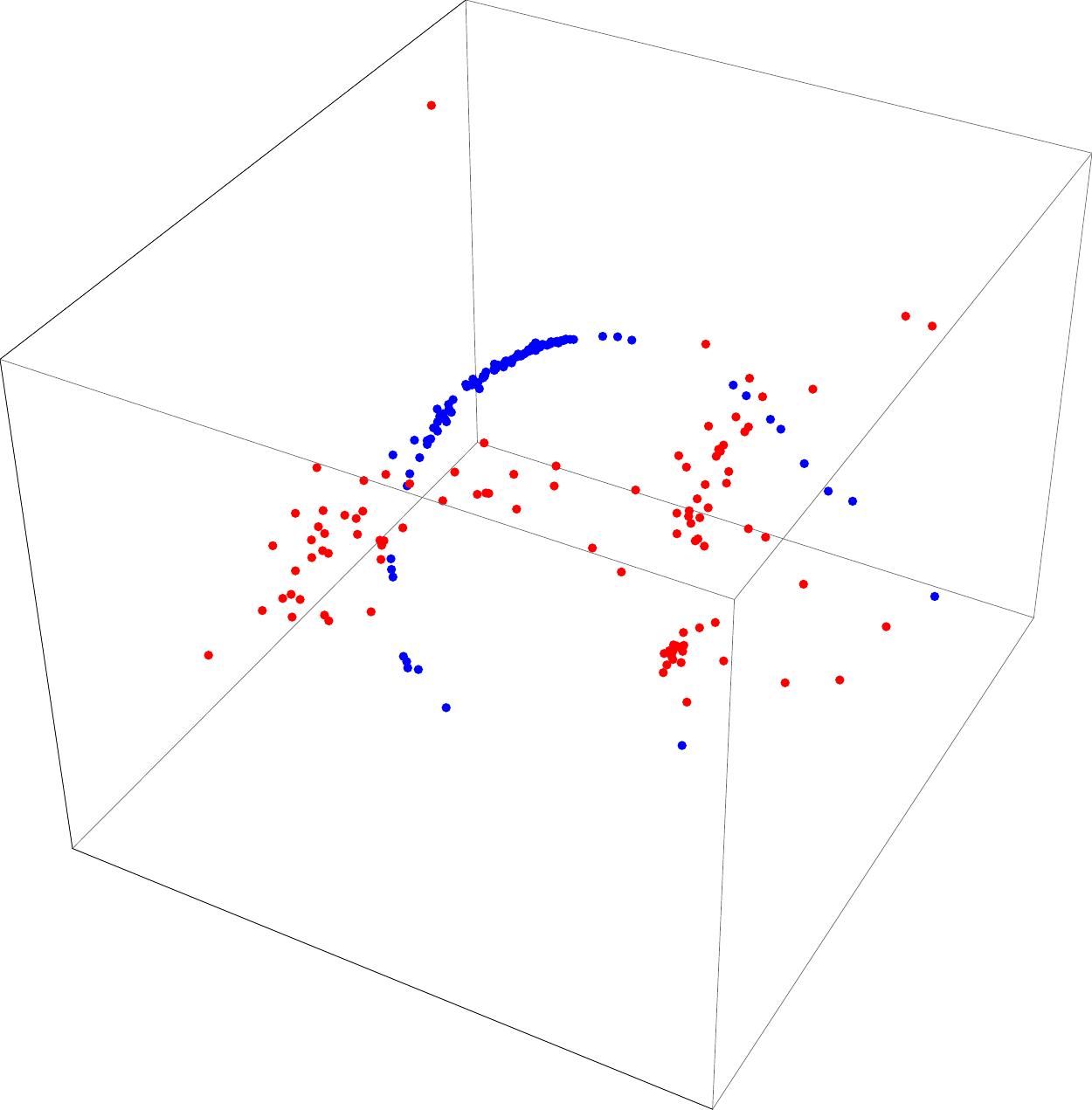}
\end{center}
\caption{Examples of ground states for 100 electric $\Gamma_e = (0,1)$ plus 100 magnetic $\Gamma_m = (1,0)$ particles.}\label{manybound}
\end{figure}
The solutions to equation (\ref{MultipBoundState}) are bound states of particles, see for example figure \ref{manybound}. Given that a system of $N$ particles has $3N$-degrees of freedom which are constrained only by $(N-1)$ equations, the bound states have a $(2N+1)$-dimensional classical moduli space $\mathcal{M}$. This moduli space cannot be accessed dynamically at low temperatures however, due to the velocity dependent forces which constrain the particles to oscillate about a fixed location if given a small kick (just like an electron in the presence of a magnetic field). As the energy is increased, the rigid structure of the bound state is deteriorated and eventually lost completely. In the following sections we study the dynamical features of this system, with a particular focus on the three-particle system. 

Equation (\ref{MultipBoundState}) is a familiar expression in supergravity. It recreates the integrability condition of~\cite{Denef:2000nb} for multi-centered black hole bound states in four-dimensional $\mathcal{N}=2$ supergravity:
\begin{equation}\label{Integrability}
\sum_{q}^N\frac{\left\langle\Gamma_p,\Gamma_q\right\rangle}{\left\lvert\mathbf{x}_p-\mathbf{x}_q\right\rvert}=2\text{Im}\left[e^{-i\alpha}Z_p\right]_{r=\infty}~.
\end{equation}
The above expression involves electric-magnetic charge vectors $\Gamma=\left(P^I,Q_I\right)$ with a duality invariant symplectic product given by:
\begin{equation}
\langle\Gamma,\tilde{\Gamma}\rangle=P^I\tilde{Q}_I-Q_I\tilde{P}^I~.
\end{equation}
Expression~(\ref{Integrability}) also involves a function called the central charge $Z_p(z)$ which depends on the vector multiplet scalars $z^a$ of the supergravity theory, as well as the charge vector $\Gamma_p$. At spatial infinity we can write $Z_p\rvert_{r=\infty}=m_pe^{i\alpha_p}$ where $m_p$ is the ADM mass of a BPS particle of charge $\Gamma_p$. If we denote $Z=\sum_pZ_p$ then the parameter $\alpha$ in~(\ref{Integrability}) is given by $\alpha=\text{arg}\left(Z\rvert_{r=\infty}\right)$. Thus we may rewrite (\ref{Integrability}) as
\begin{equation}\label{Integrability2}
\sum_{q}^N\frac{\left\langle\Gamma_p,\Gamma_q\right\rangle}{2r_{pq}}=m_p\sin\left(\alpha_p-\alpha\right)~.
\end{equation}
The supersymmetric multi-particle mechanics in (\ref{multipLag}) may be considered as a toy model for the dynamics of the multi-centered black hole bound states if we make the identifications
\begin{equation}\label{kappaeq}
\kappa_{pq}=\left\langle\Gamma_p,\Gamma_q\right\rangle\quad\text{ and }\quad\theta_p=-m_p\sin\left(\alpha_p-\alpha\right)~.
\end{equation}

\subsection{Three particles}\label{scalingsec}

Much of the discussion that follows will concern the three particle system, which already exhibits rich dynamic and non-dynamic features. Here we describe some of the characteristic features of its zero energy configurations. 

Classically, the (supersymmetric) ground states are found by setting $U_1 = U_2 = U_3= 0$. Explicitly:
\begin{equation}\label{scaling}
\frac{\kappa_{12}}{2 r_{12}} - \frac{\kappa_{31}}{2 r_{31}} = -\theta_1~, \quad -\frac{\kappa_{12}}{2 r_{12}} + \frac{\kappa_{23}}{2 r_{23}} =- \theta_2~, \quad - \frac{\kappa_{31}}{2 r_{31}} + \frac{\kappa_{23}}{2 r_{23}} = \theta_3~,
\end{equation}
with $\theta_3 = - (\theta_1 + \theta_2)$. Notice that the third equation follows from the other two. We also require that the three relative distances ${r}_{12}$, ${r}_{23}$ and ${r}_{13}$ satisfy the triangle inequality. The three particles have nine position degrees of freedom and the above equations only constrain two of them. Factoring out the center of mass leaves us with $(9-2-3)=4$ unconstrained degrees of freedom. Hence, even when three-particle bound states form, there is an infinite classical moduli space of connected (and possibly also disconnected) ground states. In the case of a two-particle bound state the classical moduli space is simply a two-sphere of fixed radius $\kappa_{12}/2\theta_1$. 

If we have $\kappa_{31}$, $\kappa_{12}$ and $\kappa_{23}$ positive we find there exist {\it scaling solutions} given by $r_{ij} \to \lambda \kappa_{ij}$ with $\lambda \to 0$ with the $\kappa$'s obeying the triangle inequality \cite{Denef:2007vg}. Thus, in this regime the particles can come arbitrarily close to each other with no cost in energy.\footnote{When thinking about scaling solutions in the gravitational context from the point of view of a far away observer, the scaling solutions are continuously connected to a single center and will look like a single centered black hole or particle. On the other hand, nearby observers will observe that the proper distance between the particles never shrinks to zero due to the formation of infinite throats (at least at zero temperature).} Away from the scaling regime the solution to (\ref{scaling}) corresponds to a bound state for which the particles may oscillate about a fixed equilibrium radius upon small perturbations. 

As an example, when $\theta_3 = -3$, $\theta_2 = -2$,  $\theta_1 = 5$, $\kappa_{12} = -1$, $\kappa_{13} = -1$ and $\kappa_{23}  = 1$, with the three particles living on a line and particle 3 between particles 2 and 1, we find the solution:
\begin{equation}
\mbox{$r_{12} = ~\frac{1}{20}\left( 7  + \sqrt{19} \right) \approx 0.57, \quad r_{13} = \frac{1}{30} \left( 8 - \sqrt{19} \right) \approx 0.12 ~, \quad r_{23} = \frac{1}{12}\left( 1 + \sqrt{19} \right) \approx 0.45~.$}
\end{equation}
Note that the triangle inequality is saturated since we have considered a collinear example. We should note, however, that there are clear instances where no solutions exist, such as when $(\kappa_{13}, \kappa_{23}, \theta_3) > (0,0,0)$, for example.

\subsection{Regime of validity}\label{sec:regvalidity}

Since we are free to choose the set of $\alpha_p$, (\ref{kappaeq}) does not really constrain the values of the $\theta_p$ in any way. There is, however, a restriction stemming from the requirement of the validity of the Coulomb branch description assuming our system comes out of integrating strings \cite{Denef:2002ru,Douglas:1996yp}.  The distances between particles must be smaller than the string scale, but larger than the ten dimensional Planck scale. For larger distances, the suitable description is given by the exchange of light closed strings (in which case supergravity is the reliable description). Furthermore, the velocities should be small compared to the speed of light to avoid higher derivative corrections to the non-relativistic Lagrangian (\ref{multipLag}). 

Consider first two particles with masses $m_1$ and $m_2$ and call the string length $l_s$. The restriction is found to be \cite{Denef:2002ru}: $\kappa_{12} \ll l_s \theta_1$ and $\mu \gg l_s^2 \theta_{1}^3/\kappa_{12}^2$, where $\mu \equiv m_1 m_2/(m_1 + m_2)$ is the reduced mass of two particles. Indeed, if $\kappa_{12} \gg l_s \theta_1$ the distance between the two particles in a bound state will be much larger than $l_s$
and the appropriate description becomes that of supergravity. When $\mu \ll l_s^2 \theta_1^3/\kappa_{12}^2$, the open strings between the branes become light and the appropriate description becomes that of the Higgs branch and eventually the fused D-brane system itself. For the multiparticle system, the Coulomb branch description is reliable so long as the inter-particle distances are sufficiently large that the massive strings can be reliably integrated out, i.e. $r_{ij} \gg l_s \sqrt{|\theta_i/m_i - \theta_j/m_j |}$ and sufficiently small that we remain in the substringy regime, i.e. $r_{ij} \ll l_s$. 

\section{Classical Phase Space}

Having discussed the general framework of the system under study, we now discuss some of its dynamical features, beginning with the classical phase space. Recall that the Hamiltonian of our system is given by:
\begin{equation}
H = \frac{1}{2 m_p} \sum_p \left[ (\mathbf{p}_p-\mathbf{A}_p)^2 + {U_p^2} \right] 
\end{equation}
The Hamilton equations of motion are given by: 
\begin{equation}
\mathbf{\nabla}_{\mathbf{x}_p} H = - \dot{\mathbf{{{p}}}}_p~, \quad \mathbf{\nabla}_{\mathbf{p}_p} H = \dot{\mathbf{{{x}}}}_p~.
\end{equation}
For $N$-particles we have a $3 \times 2 N = 6 N$ dimensional phase space. As manifest conserved quantities we have the energy, the center of mass momentum, and the center of mass angular momentum. 

\subsection{Two particles are integrable} 

We review the classical properties of the two particle problem in appendix~\ref{sec:twoparticles}. Recall that a classically integrable system with a $2N$-dimensional phase space has at least $N$ conserved quantities with mutually vanishing Poisson brackets. Phase space trajectories for integrable systems reside on $N$-dimensional tori. In the case of two-particles we have a $12$-dimensional phase space. There are six manifest conserved quantities given by the net momentum and angular momentum. As shown by D'Hoker and Vinet \cite{D'Hoker:1986uh,D'Hoker:1985kb}, the presence of a conserved Runge-Lenz vector (\ref{eq:twopartconservedquant}) leads to an enhanced $SO(3,1)$ symmetry. The angular momentum and Runge-Lenz vector are three-vectors, the Hamiltonian is a scalar and there exist two relations amongst the seven quantities, hence there is a total of $(3+3+2)=8$ conserved quantities. Factoring out the center of mass yields a (maximally) $\emph{super-integrable}$ system.\footnote{A super-integrable system \cite{wikisuper} is a system with a $2N$-dimensional phase space which has {\it more} than $N$ conserved quantities. A maximally super-integrable system has $2N-1$ conserved quantities.} The super-integrability implies its equations can be separated in more than one coordinate system and one can solve for the quantum mechanical spectrum algebraically, as done in \cite{Avery:2007xf}. Further, it implies that trajectories in coordinate space follow paths which are closed in the case of bound orbits. 


\subsection{Three particles are chaotic}

In the case of three-particles we have an $18$-dimensional phase space and there is no longer a sufficient number of conserved quantities to render the system integrable. Thus, such systems will exhibit chaotic behavior. 

We may use several numerical tools to analyze the chaotic nature of such a system. For instance, we can study the Lyapunov exponent $\lambda$ parameterizing the divergence of phase space trajectories with nearby initial conditions. Given two trajectories in phase space with initial separation $\delta z_0$, the Lyapunov exponent is defined by the limit:
\begin{equation}
\lambda = \lim_{t \to \infty} \lim_{\delta z_0 \to 0} \frac{1}{t} \log \frac{\delta z(t)}{\delta z_0}~.
\end{equation}

We can also study Poincar\'{e} sections in phase space. These are found by recording the location of a trajectory in a particular subspace of phase space each time it crosses some fiducial point (such as crossing the origin with positive velocity). These are particularly useful for lower dimensional systems such as the double pendulum, where they clearly depict the breakdown of the integrable motion on a two-torus as the energy  is increased (see chapter 11 of \cite{hand} for a discussion). We will discuss and examine the Poincar\'{e} sections of a collinear three particle system in section \ref{doublepend}.


Six of the phase space dimensions can be eliminated from net momentum conservation and factoring out the center of mass. We can also kill another four due to net angular momentum and energy conservation. The remaining $8$-dimensions in phase space (as far as we know) are unconstrained by symmetries. Needless to say, systems with more than three particles will also display chaotic properties. A simpler setup, which we refer to as the Euler-Jacobi setup, is that of a probe particle moving in the background of two fixed centers. 


\section{Euler-Jacobi Ground States}

The simplest question we can ask about our system is what the (supersymmetric) ground states are, both classically and quantum mechanically. Classically there may be continuous moduli spaces of zero energy configurations. Quantum mechanically, given that the probe is a particle in the presence of a background magnetic field, we expect the continuous classical moduli space to give rise to a degenerate set of quantum ground states due to Landau degeneracies. 

\subsection{Euler-Jacobi three body problem}

We will consider a probe particle of mass $m_3$ in the background of two fixed centers unless otherwise specified. The background particles have masses $m_1$ and $m_2$ both very large compared to $m_3$, charge vectors $\Gamma_1$ and $\Gamma_2$ with symplectic product $\kappa_{12} > 0$ and Fayet-Iliopoulos constant $\theta_1 = - \theta_2 < 0$. They sit along the $z$-axis at $z  = \pm {\kappa_{12} }/{4\theta_1} \equiv \pm{a}$. By choosing $m_1$ and $m_2$ much larger than $m_3$, the backreaction of the probe on the fixed centers is suppressed by $\mathcal{O}(m_3/m_1,m_3/m_2)$. The probe also has charge vector $\gamma_3$ and Fayet-Iliopoulos constant $\theta_3$. 

We may also consider the possibility  of forming supersymmetric bound states between the probe and the fixed centers. In such a case, a non-zero $\theta_3$ requires us to modify the background condition $\theta_1 = -\theta_2$, since now the $\theta$'s must add to zero. We thus demand $|\theta_3| \ll |\theta_1|, |\theta_2|$ such that the correction to the positions of the original fixed centers is of order $\mathcal{O}(\theta_3/\theta_1,\theta_3/\theta_2)$. The intersection products of the probe with the centers are given by $\kappa_{31}$ and $\kappa_{32}$. To avoid any large backreaction from the probe due to the $\kappa$ interactions we further require that $\kappa_{31}/r_{31}$ and $\kappa_{32}/r_{32}$ are small compared to $\kappa_{12}/r_{12} \sim \theta_2$.


The Hamiltonian governing the dynamics of the probe is given by:
\begin{equation}\label{hprobe}
H_{probe} = \frac{1}{2 m_3} \left( \textbf{p}_3 - \textbf{A}_3 \right)^2 + \frac{1}{2m_3} \left( \theta_3 + \frac{\kappa_{31}}{2 r_{31}} +  \frac{\kappa_{32}}{2 r_{32}}\right)^2~.
\end{equation}
Notice that the two scaling transformations:
\begin{equation}
(\textbf{p}_3(t),\textbf{x}_3(t) \; ; \; \kappa_{ij},\theta_3,m_3,r_{12},t) \to (\sigma \; \textbf{p}_3(t),\lambda \; \sigma^{-1} \; \textbf{x}_3(t) \; ; \; \lambda \; \kappa_{ij},\sigma \; \theta_3, \sigma^2 \; m_3, \lambda \; \sigma^{-1} \;  r_{12},\lambda \; t)~,
\end{equation}
generate a family of solutions parameterized by $\lambda$ and $\sigma$. Given a solution to the equation of motion for some $\mathcal{O}(1)$ parameters, we can exploit the scaling symmetries to map the solution to a rescaled one in the regime of validity for the Coulomb branch description as discussed in section \ref{sec:regvalidity}. In particular, we require $\lambda \gg \sqrt{\sigma}$ and $\lambda \ll \sigma$ which can be achieved for large $\sigma$. Notice that in this regime the velocity (which scales as $\sigma^{-1}$) becomes parametrically small. 



\subsection{Classical Ground States}

As noted in (\ref{MultipBoundState}), the space of classical ground states $\mathcal{M}$ is given by setting $U_p = 0$. Satisfying this condition gives rise to time independent classical bound states. In the probe limit, where the two background centers are fixed, this amounts to solving the algebraic equation:
\begin{equation}
\frac{1}{2}\frac{\kappa_{31}}{\sqrt{\rho^2 + (z-a)^2}} + \frac{1}{2} \frac{\kappa_{32}}{\sqrt{\rho^2 + (z+a)^2}} =  - \theta_3~,
\end{equation}
where $\rho^2 = x^2 + y^2$ and $\phi = \tan^{-1} (y/x)$. One can easily prove that the effective magnetic field $\textbf{B} = \boldsymbol{\nabla}_3 \times \textbf{A}_3$ is always perpendicular to the tangent of $\mathcal{M}$. This remains true for the moduli space of a probe in a background of more than two centers as well.

Consider first the case $\theta_3 = 0$. We find:
\begin{equation}\label{surface}
\rho^2 = \frac{{(a-z)^2 \kappa_{32}^2-(a+z)^2 \kappa_{31}^2}}{{(\kappa_{31}^2 - \kappa_{32}^2)}}~.
\end{equation}
For $\rho(z)$ above to have solutions we choose $\kappa_{31} > 0 > \kappa_{32}$ and furthermore $|\kappa_{31}| > |\kappa_{32}|$, we find a continuum of solutions between $z = [z_-,z_+]$ where:
\begin{equation}
z_\pm = \pm a  \frac{|\kappa_{32}| \mp |\kappa_{31}| }{|\kappa_{31}| \pm |\kappa_{32}|} ~.
\end{equation}
Note that $z_- < -a < z_+ < a$ and thus the $\theta_3 = 0$ surfaces enclose the fixed charge at $z = -a$ in this case. Similarly, for $|\kappa_{31}| < |\kappa_{32}|$ the probe encloses only the center at $z=a$. For $\kappa_{31} = - \kappa_{32}$ and $\theta_3$, the classical moduli space $\mathcal{M}$ becomes the $z = 0$ plane.

\begin{figure}
\begin{center}
{\includegraphics[height=70mm]{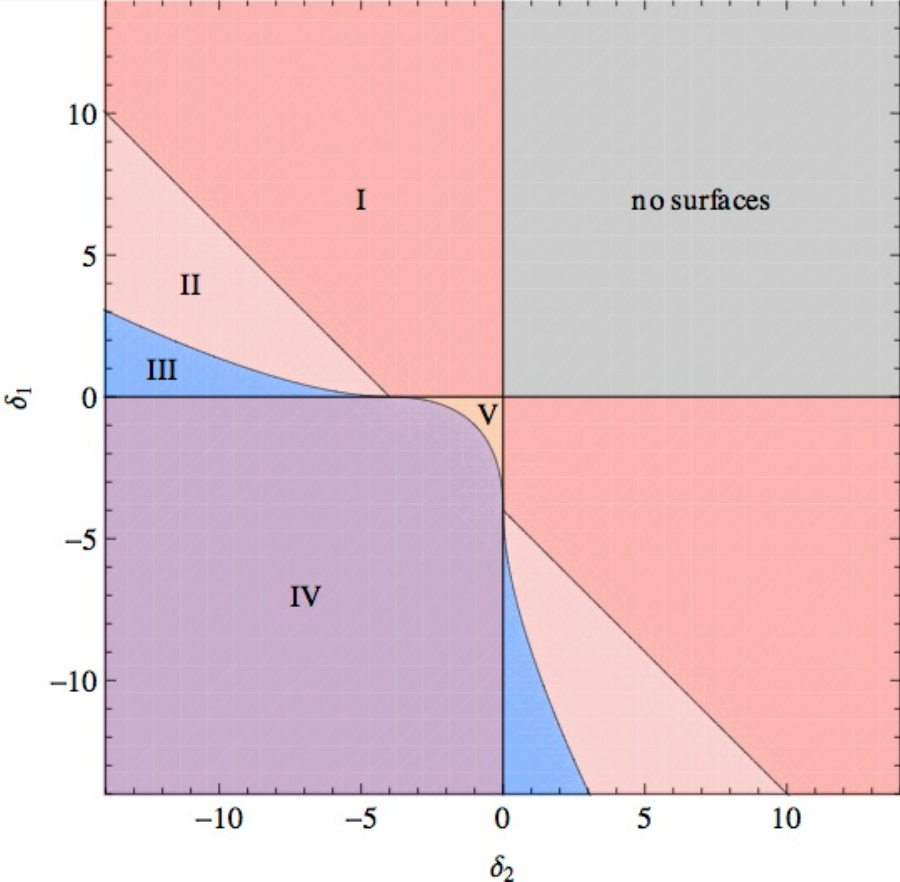}~\quad \includegraphics[height=70mm]{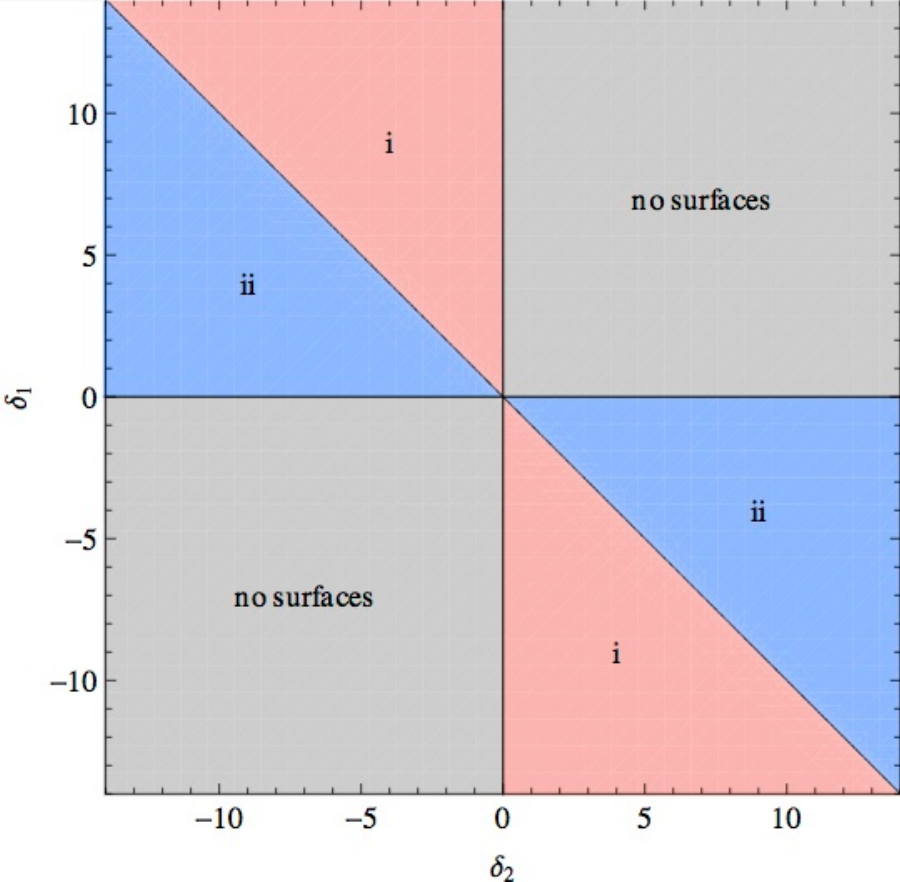}}
\end{center}
\caption{Left: Classical moduli space $\mathcal{M}$ in the $\delta_1 - \delta_2$ plane for $\theta_3 \neq 0$. The nature of $\mathcal{M}$ for the different regions is shown in figure \ref{regions}. Right: Classical moduli space for $\mathcal{M}$ with $\theta_3 = 0$. In regions i and ii the centers at $z = a$ and $z = -a$ are enclosed respectively.}\label{phaseplot}
\end{figure}

\begin{figure}
\begin{center}{
\subfloat[position=top][Region I]{\includegraphics[height=50mm]{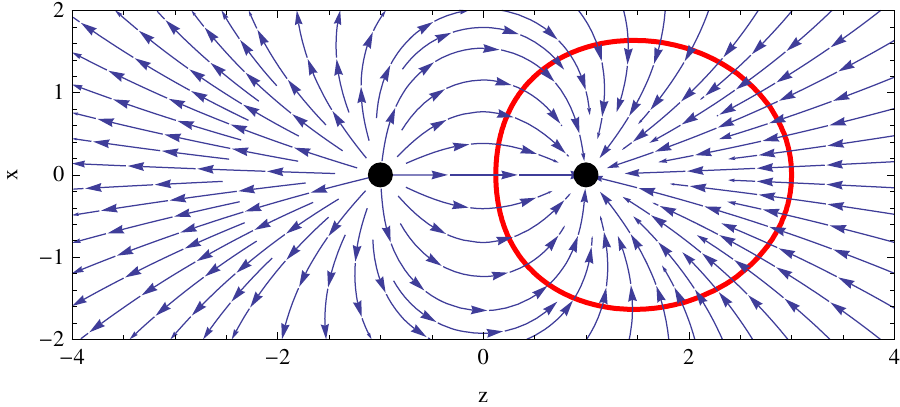}} \\
\subfloat[Region II]{\includegraphics[height=50mm]{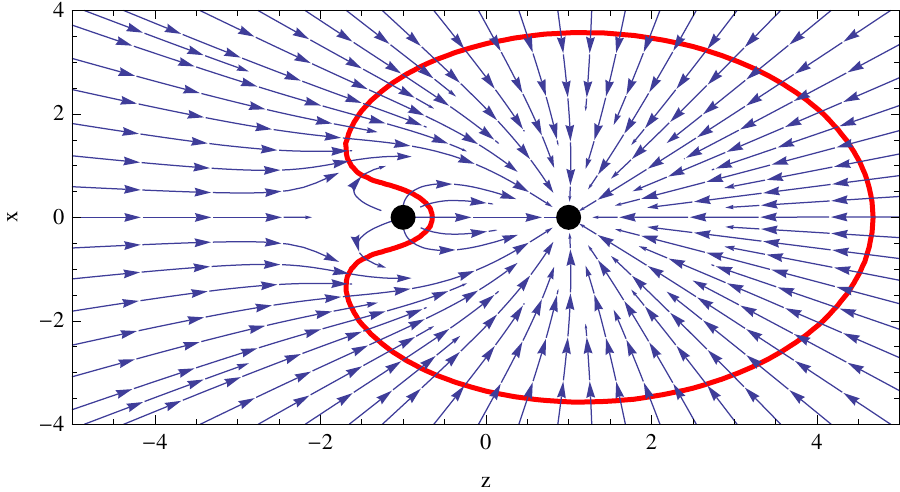}} 
\subfloat[Region III]{\includegraphics[height=50mm]{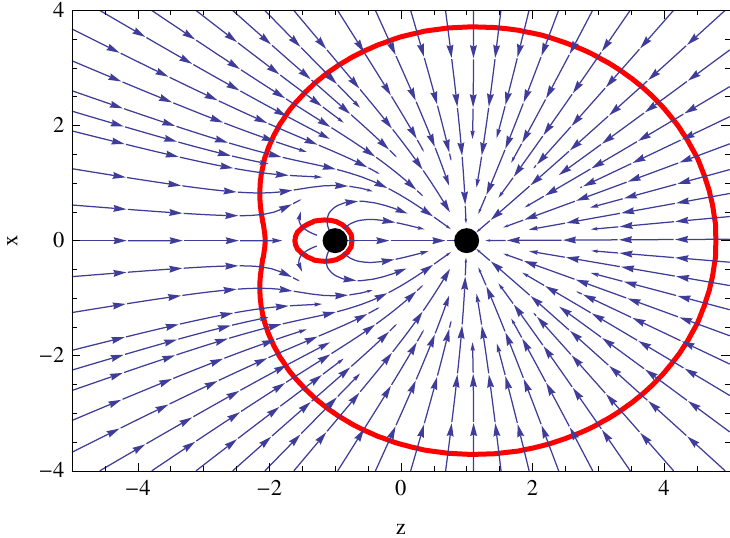}}  \\ 
\subfloat[Region IV]{\includegraphics[height=50mm]{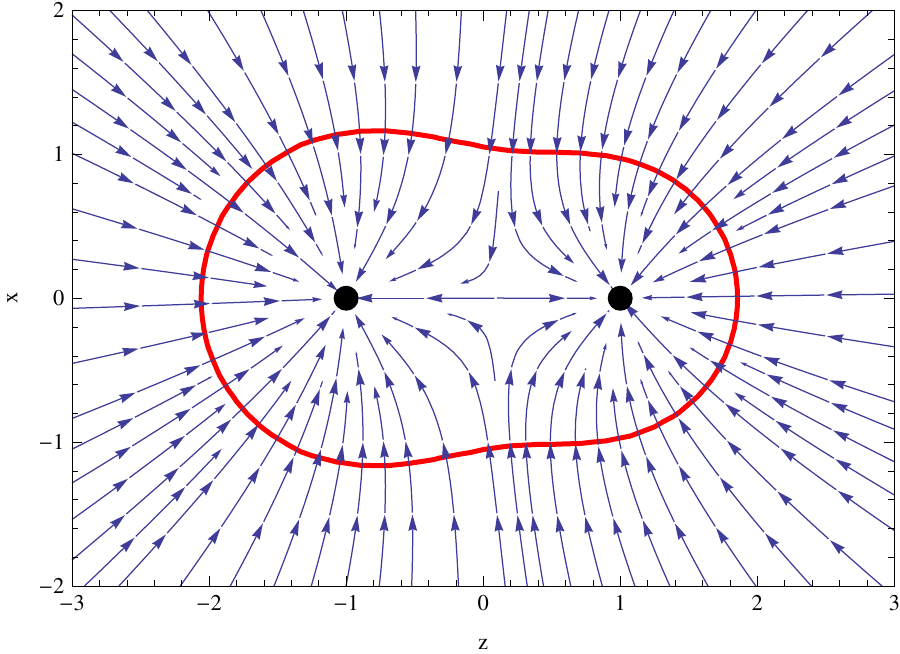}} 
\subfloat[Region V]{\includegraphics[height=50mm]{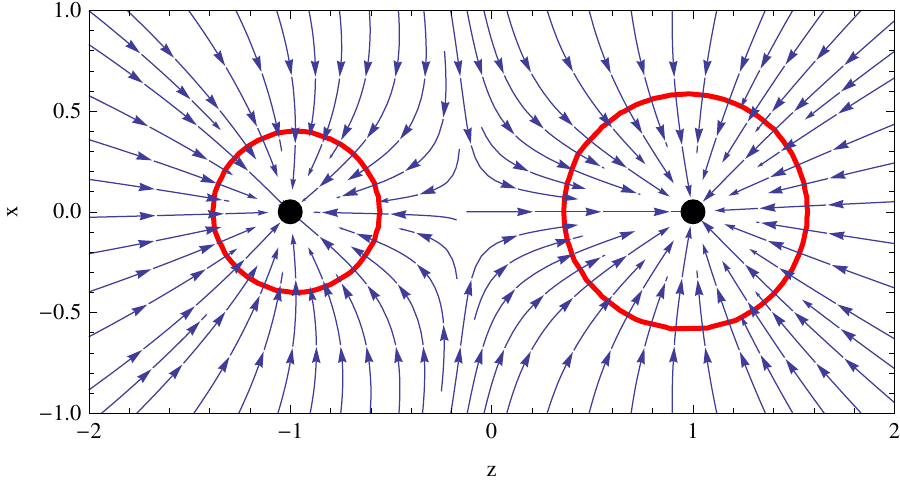}}
}
\end{center}
\caption{Classical moduli space in the $\delta_1 - \delta_2$ plane for $\theta_3 \neq 0$. The order of the figures left to right starting at the top are the regions in figure \ref{phaseplot}.}\label{regions}
\end{figure}

For $\theta_3 \neq 0$ finding $\rho(z)$ amounts to solving a quartic equation. In order to do so, it is convenient to go to prolate spheroidal coordinates:
\begin{equation}\label{prolate}
\rho = a \sqrt{(\xi^2-1)(1-\eta^2)}~, \quad z = a \xi \eta~, \quad \phi = \phi~,
\end{equation}
such that:
\begin{equation}\label{u30eqetaxi}
2 a \theta_3 (\eta^2-\xi^2) = (\kappa_{31}+\kappa_{32})\xi+(\kappa_{31}-\kappa_{32})\eta~.
\end{equation} 
In the above we have implicitly used that $\eta\in[-1,1]$ and $\xi\in[1,\infty]$. We can easily find a solution for $\eta = \eta(\xi)$:
\begin{equation}\label{u30eqetaxisol}
\eta(\xi) =\frac{1}{4}\left(\delta_{1}-\delta_{2}\pm\sqrt{(\delta_1-\delta_2)^2+8(\delta_1+\delta_2)\xi+16\xi^2}\right)~,
\end{equation}
where $\delta_1 \equiv \kappa_{31}/(a \theta_3)$ and $\delta_2 \equiv \kappa_{32}/(a \theta_3)$. In figure \ref{phaseplot} we show the different qualitative types of $\mathcal{M}$ as a function of $\delta_1$ and $\delta_2$. The qualitative features of each region are shown in figure \ref{regions}. Notice that upon defining the prolate coordinates (\ref{prolate}) we have scaled out the distance $r_{12} = a$ between the fixed centers. To obtain physical distances we simply multiply by $r_{12} = a$.



\subsection{Quantum Ground States}

From the classical point of view, our particle is nothing more than a charged particle in the presence of magnetic fields constrained to live on a surface. Thus, given a time independent supersymmetric bound state configuration we can compute the lowest Landau degeneracies $d_L$ by computing the degeneracy of states with vanishing energy for the constrained particle. Such a setup has been addressed for non-uniform magnetic fields everywhere normal to the surface \cite{alicki}, which is precisely the situation we find ourselves in. Following \cite{Denef:2002ru}, the lowest Landau degeneracies are given by the total flux through the classical moduli space $\mathcal{M}$. For instance, as we discuss below, the Landau degeneracy of the fixed background is given by $\kappa_{21}$. Upon studying the phase diagram and corresponding $\mathcal{M}$ in figures \ref{phaseplot} and \ref{regions}, we find that the total degeneracy is:
\begin{eqnarray}
\text{I, II and III} &:&  d_{tot} = \kappa_{12} \times | \kappa_{31} |\quad\text{or}\quad d_{tot} = \kappa_{12} \times | \kappa_{32} |~,\\
\text{IV and V} &:& d_{tot} = \kappa_{12} \times | \kappa_{31} + \kappa_{32} |~. 
\end{eqnarray}
For regions I, II and III, the degeneracy of states depends on which of the two background centers is encircled by $\mathcal{M}$. Notice there is a jump in the number of ground states as we vary $\delta_1$ and $\delta_2$. Since we are in the probe limit, we expect these results to be correct up to order $\mathcal{O} (\kappa_{31}/\kappa_{12})$ and $\mathcal{O} ( \kappa_{32}/\kappa_{12} )$.

From the supersymmetric quantum mechanics point of view, recall that $\{Q_\alpha,\bar{Q}^\beta \} = 2 \delta_\alpha^\beta H $.  In the absence of the probe, the ground state of the background is given by \cite{Denef:2002ru}:
\begin{equation}
| \textbf{b} \rangle = \Psi_\alpha (\vec{x}_1 - \vec{x}_2) \bar{\tilde{\lambda}}^\alpha | 0 \rangle~, \quad \tilde{\lambda} \equiv \lambda_1 - \lambda_2~.
\end{equation}
The center of mass coordinate $\vec{x}_0 \equiv \left( m_1 \vec{x}_1 + m_2 \vec{x}_2 \right)/(m_1 + m_2)$ and center of mass spinor $\lambda_0 \equiv \left( m_1 \lambda_1 + m_2 \lambda_2 \right)/(m_1 + m_2)$ drop out and thus $| \textbf{b} \rangle$ is naturally a function of the relative background position vector and spinor. The state $| 0 \rangle$ is annihilated by $\tilde{\lambda}$ and defines a three-dimensional Hilbert space through action of $\bar{\tilde{\lambda}}$. There are $\kappa_{12}$ ground states filling a spin-$(\kappa_{12}-1)/2$ multiplet.
From the last term in the Hamiltonian (\ref{ham2}), we observe that there exist spin-spin couplings between the background spinors $\lambda_1$ and $\lambda_2$ and the probe spinor $\lambda_3$. It is convenient to introduce the relative spinors $\lambda_{13} \equiv \lambda_{1}-\lambda_{3}$ and $\lambda_{23} \equiv \lambda_2 - \lambda_3$ and their associated vacua $| 0_{23} \rangle$ and $| 0_{13} \rangle$, such that $\lambda_{23} | 0_{23} \rangle = 0$ and so on. Both  $| 0_{23} \rangle$ and  $| 0_{13} \rangle$ have an associated three-dimensional Hilbert space and the general state must be a tensor product of all linear combinations of all such states, finally tensored with $| \textbf{b} \rangle$. Given that the probe is sensitive to the background $\textbf{B}$-field, it will go into spin one-half states of $|0_{13}\rangle$ and $|0_{23}\rangle$ aligning with the $\textbf{B}$-fields from the fixed particles at $z = \pm a$. This will split the lowest Landau degeneracies. It would be interesting to compute the explicit ground state wavefunctions.

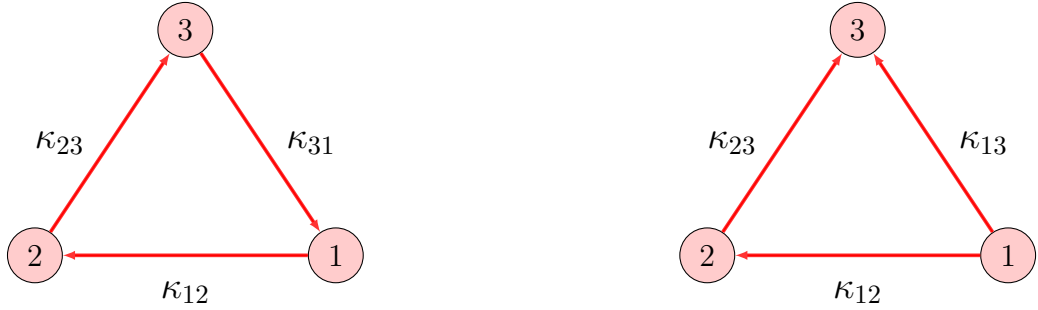
\begin{figure}

\begin{minipage}[b]{0.5\linewidth}
\centering
\begin{tikzpicture}
 \tikzset{LabelStyle/.style =   {                                text           = black}}

\tikzstyle{line} = [draw=red, -latex']
  \node[fill=red!20,draw,circle] (n2) at (0,0) {2};
  \node[fill=red!20,draw,circle]  (n1) at (4,0)  {1};
  \node[fill=red!20,draw,circle]  (n3) at (2,3)  {3};

 \draw[->, >=latex, color=red!80,double=red] (n1) to node[LabelStyle,below=.2cm]{{\large $\kappa_{12}$}} (n2) ;
 \draw[->, >=latex, color=red!80,double=red](n2) to node[LabelStyle,left=.2cm]{{\large$\kappa_{23}$}} (n3);
 \draw[->, >=latex, color=red!80,double=red](n3) to node[LabelStyle,right=.2cm]{{\large$\kappa_{31}$}} (n1);
\end{tikzpicture}
\label{fig:figure1}
\end{minipage}
\hspace{0.5cm}
\begin{minipage}[b]{0.5\linewidth}
\centering
\begin{tikzpicture}
 \tikzset{LabelStyle/.style =   {                                text           = black}}

\tikzstyle{line} = [draw=red, -latex']
  \node[fill=red!20,draw,circle] (n2) at (0,0) {2};
  \node[fill=red!20,draw,circle]  (n1) at (4,0)  {1};
  \node[fill=red!20,draw,circle]  (n3) at (2,3)  {3};

 \draw[->, >=latex, color=red!80,double=red] (n1) to node[LabelStyle,below=.2cm]{{\large $\kappa_{12}$}} (n2) ;
 \draw[->, >=latex, color=red!80,double=red](n2) to node[LabelStyle,left=.2cm]{{\large$\kappa_{23}$}} (n3);
 \draw[->, >=latex, color=red!80,double=red](n1) to node[LabelStyle,right=.2cm]{{\large$\kappa_{13}$}} (n3);
\end{tikzpicture}
\label{fig:figure2}
\end{minipage}
\caption{Three node quiver with a closed loop (left) and without a closed loop (right).}\label{quiverdiag}
\end{figure}

For the sake of completeness we briefly mention another method to compute the number of ground states. One can associate a quiver diagram $Q$ to the data $(\kappa_{ij},\theta_i)$ of a particular configuration \cite{Douglas:1996sw,Denef:2002ru}. It turns out that the dimension of the moduli space $\mathcal{M} ( Q,\textbf{N},\theta)$ of the quiver $Q$ can be related to the number of BPS ground states. In particular, for the three body problem where each particle is a different species we have a quiver theory Q with $\textbf{N} = (1,1,1)$, $\kappa_{12}$ arrows between nodes $1$ and $2$, $\kappa_{13}$ arrows between nodes $1$ and $3$ and $\kappa_{23}$ arrows between nodes $2$ and $3$. The quiver diagram is presented in figure \ref{quiverdiag}. The Fayet-Iliopoulos constants $\theta_v$ are additional parameters associated with each node $v$. Ground state degeneracies for similar setups to the one we are studying have been computed in \cite{Denef:2007vg}. Notice that scaling solutions can occur only for quivers with closed loops. In our problem, with $\kappa_{12} > 0$, we find that regions I, II and III correspond to quivers with closed loops and regions IV and V correspond to quivers with no closed loops.

\section{Euler-Jacobi Dynamics: classical integrability}\label{integrable}


The equations governing the probe are dictated by the Hamiltonian in (\ref{hamiltonian}). There are two obvious constants of motion in this problem, namely the energy and the angular momentum in the direction of the line where the two centers are placed. If the system is to be rendered integrable, there must exist a third constant of motion. Such a constant of motion was found for the problem of a Newtonian probe interacting gravitationally with a background of two fixed massive particles \cite{twocenter,Eulermemoir}, also known as the Euler-Jacobi three-body problem. We will show that the analogous problem in the theory under consideration is also integrable. This was previously shown and discussed in \cite{Nersessian:2007gc, Krivonos:2006qd, Bellucci:2008rp, Bellucci:2007kx}.

\subsection{Setup and coordinate systems}

Recall that we are considering two fixed background centers sitting on the $z$-axis at $z = \pm {\kappa_{12} } / 4a{\theta_1} \equiv \pm{a}$. Let us go to a cylindrical system with metric:
\begin{equation}
ds^2 = d\rho^2 + \rho^2 d\phi^2 + dz^2~,
\end{equation}
where the Cartesian and cylindrical coordinates are related by $x= \rho \cos \phi$,  $y = \rho \sin \phi$ and $z = z$. One observes that the Lagrangian and Hamiltonian are independent of the $\phi$ coordinate which implies a symmetry. The conserved quantity of this symmetry is given by the angular momentum in the $z$-direction, such that the canonical momentum $p_{\phi} = l$ is constant. The probe Hamiltonian (\ref{hprobe}) in cylindrical coordinates becomes:
\begin{equation}
H_{probe} = \frac{1}{2 m_3} \left(  p'_i - A_i (\rho,\phi,z) \right)g^{ij}  \left(  p'_j - A_j (\rho,\phi,z)\right) + \frac{\left(U_3(\rho,z)\right)^2}{2m_3}~, 
\end{equation} 
where the $p'_i$ are the conjugate momenta in the cylindrical coordinates. The relation between conjugate momenta between the primed and unprimed coordinate systems is $p_i = p'_j \; \partial x'^{j}/\partial x^i$. 

\begin{figure}
\begin{center}
{\includegraphics[height=50mm]{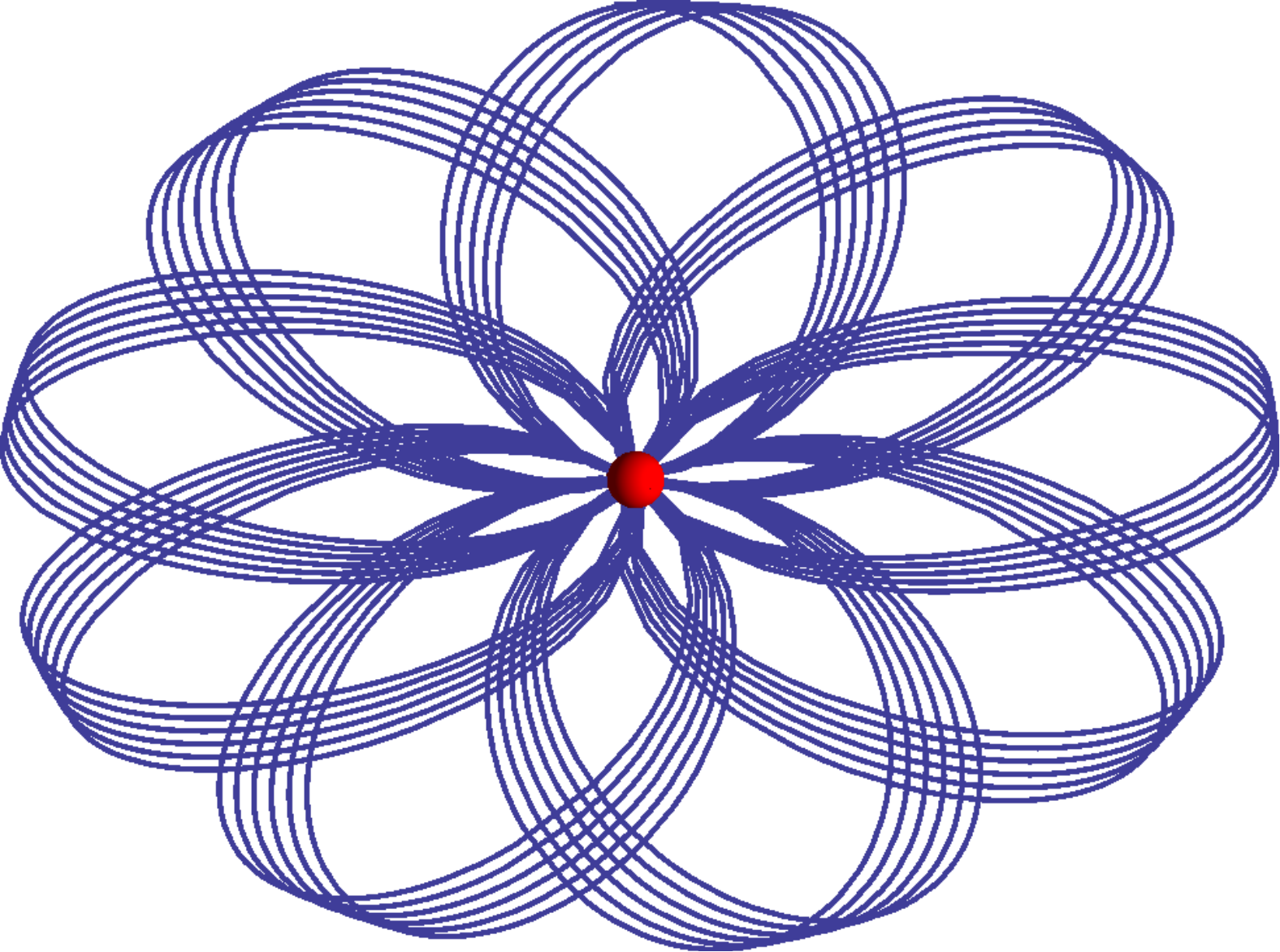} \hspace{28mm} \includegraphics[height=50mm]{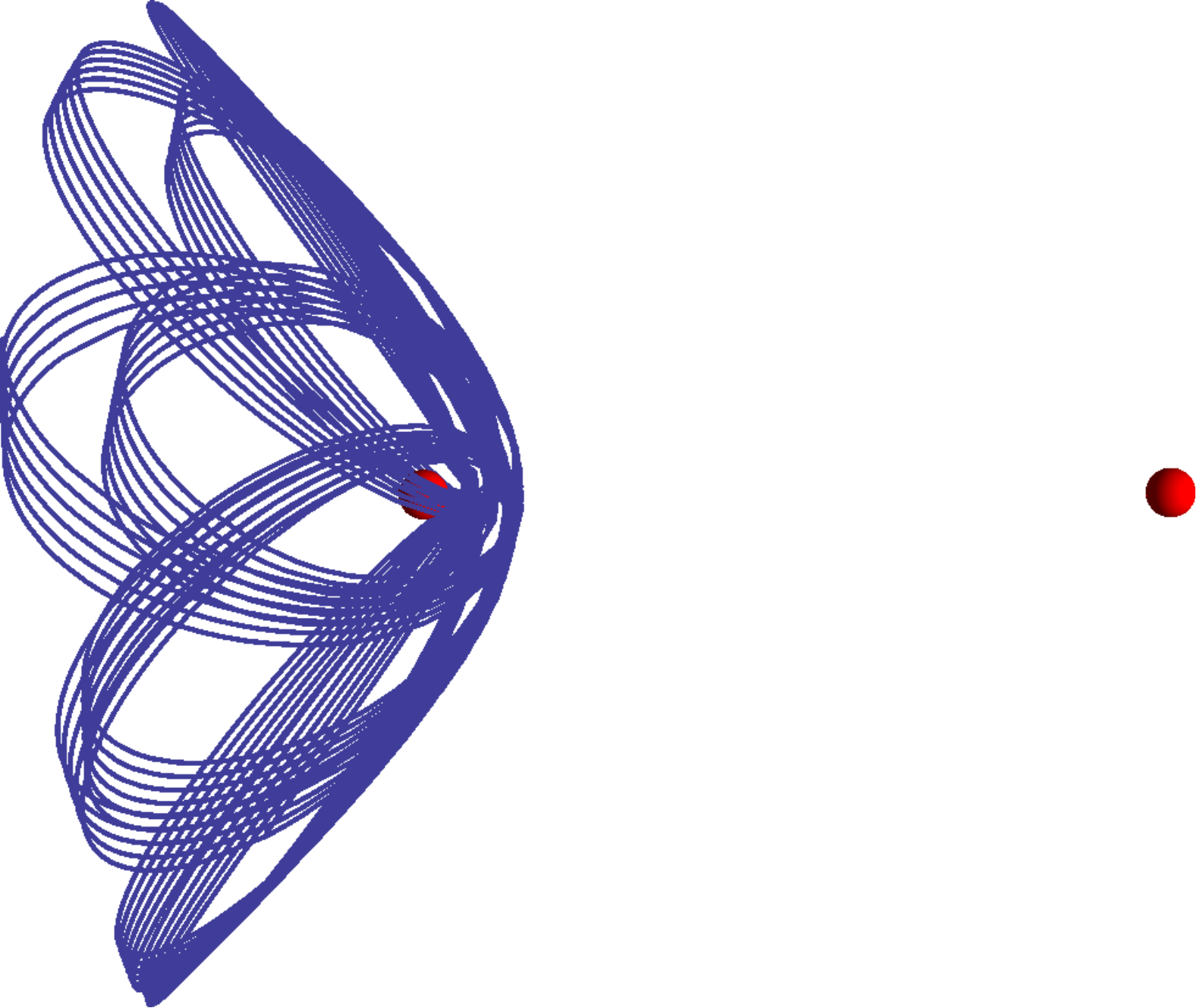}}
\end{center}
\caption{The Euler-Jacobi flower. The red balls represent the fixed background centers and the blue line represents the classical trajectory of the probe. In this case, the trajectory precesses around only one of the fixed centers.}
\end{figure}\label{closed}

The third constant of motion is not manifest in cylindrical coordinates. One must go to the prolate spheroidal coordinates (\ref{prolate})
with metric:
\begin{equation}
ds^2 = a^2 (\xi^2 - \eta^2) \left[ \frac{d\xi^2}{(\xi^2 - 1)}  +  \frac{d\eta^2}{(1 - \eta^2)}  \right] + a^2 (\xi^2-1)(1-\eta^2) d\phi^2~.
\end{equation}
Once in this coordinate system, we note that our Hamiltonian takes the following form:
\begin{equation}
H_{probe} = \frac{H_\xi + H_\eta}{\xi^2 - \eta^2}~,
\end{equation}
where $H_\xi$ depends {\it only} on $\xi$ and $p_\xi^2$ and $H_\eta$ depends {\it only} on $\eta$ and $p_\eta^2$. Thus, we can write:
\begin{equation}
H_{probe} \xi^2 - H_\xi = H_\eta + H_{probe} \eta^2 \equiv G~,
\end{equation}
where $G$ must be a constant of motion. More explicitly we have:
\begin{equation*}
H_\xi = p_\xi^2 \frac{(\xi^2-1)}{2 a^2 m_3} +  \frac{p_\phi^2}{2 a^2 m_3(\xi^2-1)} + p_\phi \frac{ \xi(\kappa_{31}-\kappa_{32}) }{2 a^2 m_3(\xi^2-1)}
+
\frac{(\kappa_{31}-\kappa_{32})^2}{8 a^2 m_3(\xi^2-1)}
+
\frac{\theta_3 \xi(\kappa_{31}+\kappa_{32}+a\theta_3 \xi)}{2am_3}
\label{hamiltonian_xi}~,
\end{equation*}
and
\begin{equation*}
H_\eta=
p_\eta^2\frac{(1-\eta^2)}{2 a^2 m_3}
+
 \frac{p_\phi^2}{2 a^2 m_3(1-\eta^2)}
+
p_\phi \frac{ \eta(\kappa_{31}+\kappa_{32}) }{2 a^2 m_3(\eta^2-1)}
+
\frac{(\kappa_{31}+\kappa_{32})^2}{8 a^2 m_3(1-\eta^2)}
+
\frac{\theta_3 \eta(\kappa_{31}-\kappa_{32}-a\theta_3 \eta)}{2 a m_3}
\label{hamiltonian_eta}~.
\end{equation*}
We conclude that the probe-two-center problem of supergoop is integrable, providing another example to the distinguished list of integrable classical systems. In this system, one observes highly symmetric spatial trajectories, as illustrated in figure \ref{closed}.

\section{Beyond Euler-Jacobi: the stringy double pendulum}\label{doublepend}

\begin{figure}
\begin{center}
{\includegraphics[height=60mm]{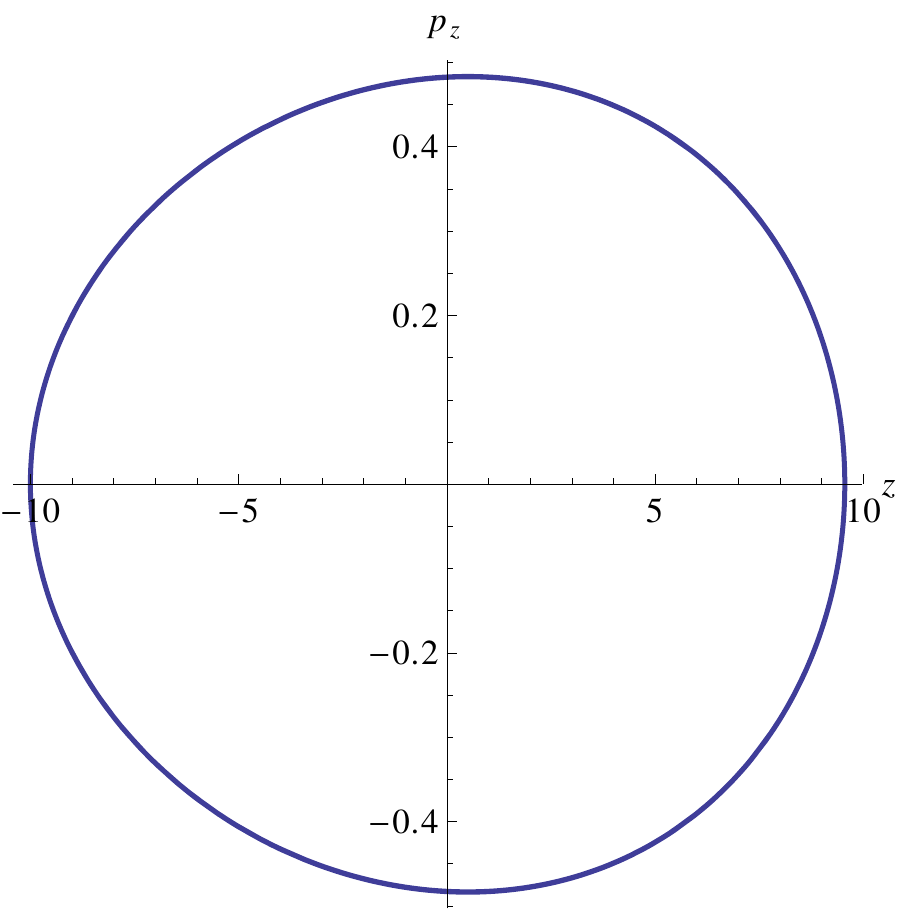} \quad \quad \quad \quad \includegraphics[height=60mm]{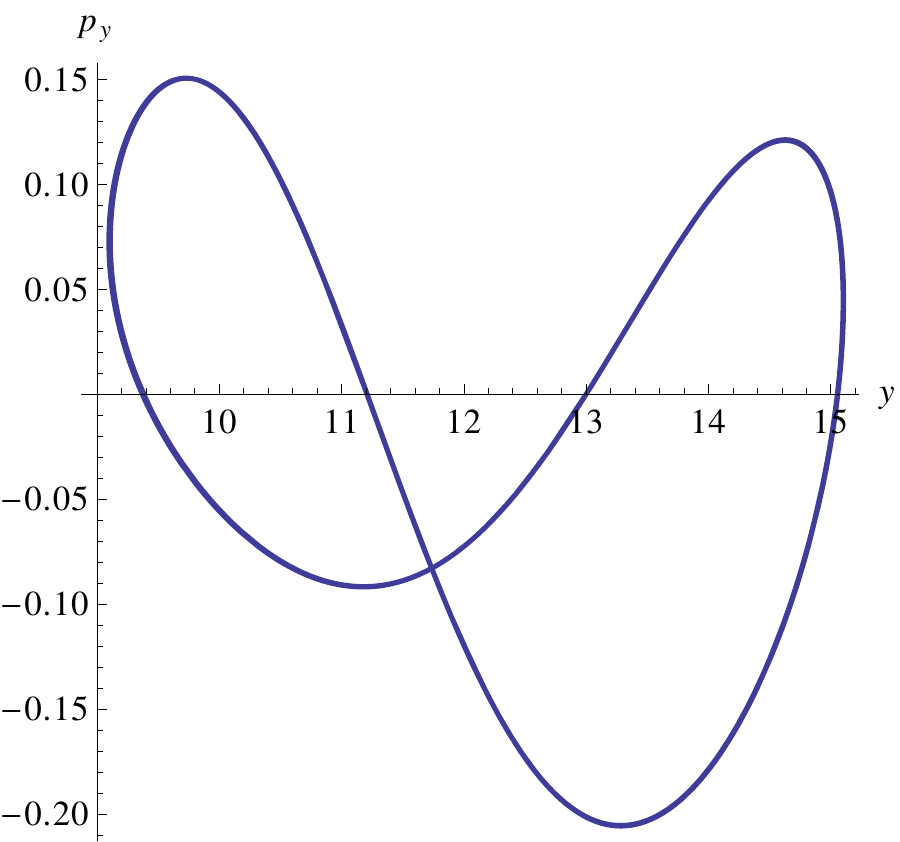}}
\end{center}
\caption{Examples of closed phase space trajectories in the integrable probe regime. The plots show slices of phase space in the Cartesian coordinate system.} 
\label{probe}
\end{figure}

\begin{figure}
\begin{center}
\includegraphics[height=60mm]{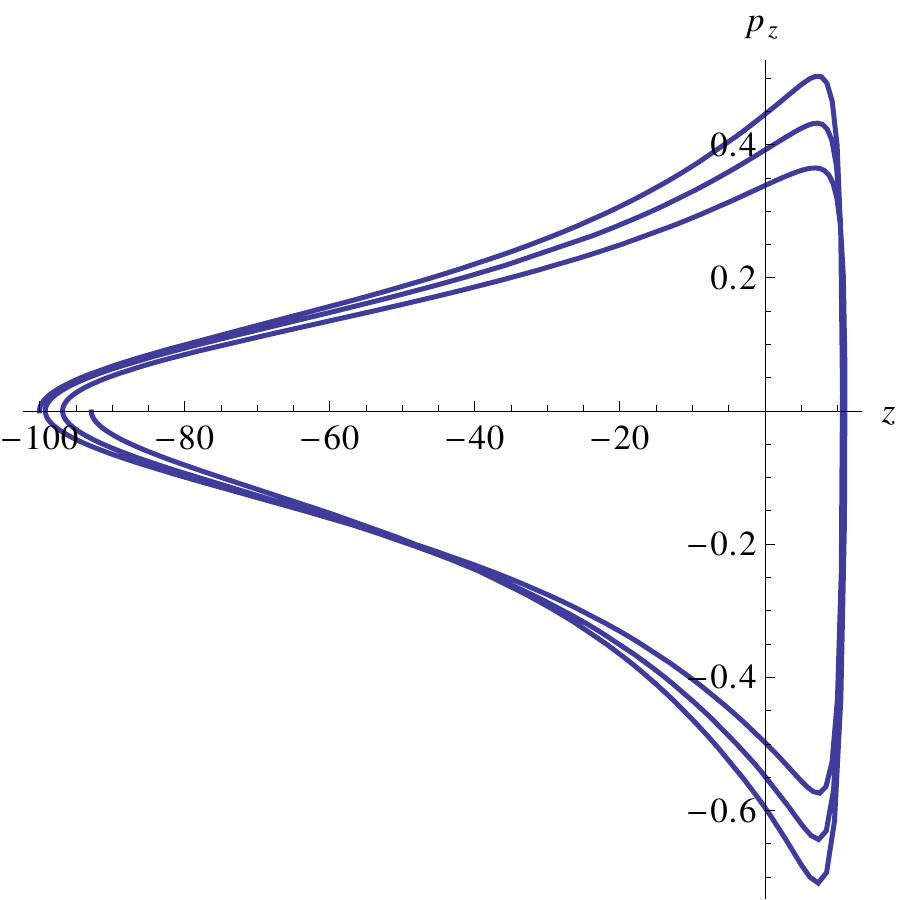} \quad \quad \quad \quad \includegraphics[height=60mm]{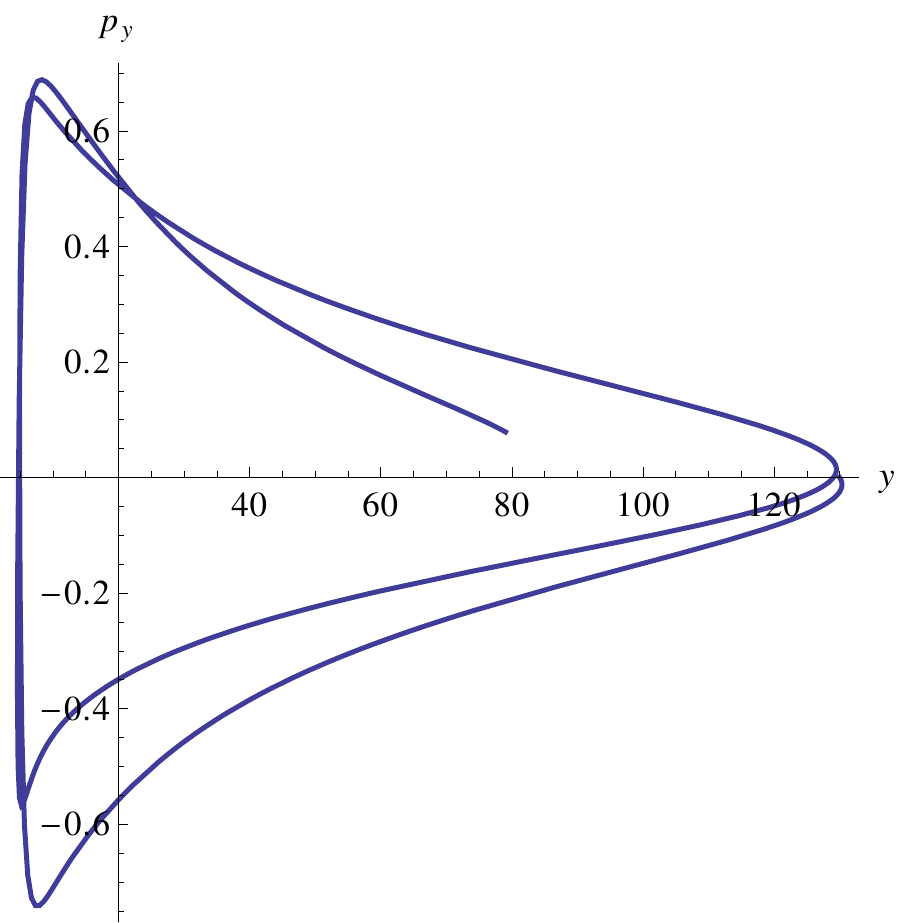} 
\end{center}
\caption{Examples of open phase space trajectories in the chaotic regime. The plots show slices of phase space in the Cartesian coordinate system.}\label{outofprobe}
\end{figure}

If we move away from the probe approximation and allow backreaction with the fixed centers, our system is no longer integrable and begins to show chaotic features. For instance, one can study trajectories in phase space and see whether they are closed. One could also compute the Lyapunov coefficient of the system. In figures \ref{probe} and \ref{outofprobe} we demonstrate the trajectories in phase space for the probe orbiting around both centers as we exit the probe limit.

As we increase the number of degrees of freedom, the analysis of chaotic systems becomes increasingly challenging. The canonical example of a double pendulum, which already displays a significant set of features generic to chaotic systems at large, can be effectively analyzed with the use of Poincar\'{e} sections. For a double pendulum, the phase space is four-dimensional with a single constant of motion -- the total energy. Hence the system is not integrable. 

\subsection{Collinear dynamics}

Away from the probe limit, as we already noted, our system has an 18-dimensional phase space and becomes a complicated three-body problem. In order to study the transition to chaos of our system, it is instructive to find a setup that allows us to use the same tools used to analyze the double pendulum. This can be achieved by restricting the particles to be collinear, i.e. placing them on a line and only considering deformations along this direction. Notice that a system consisting of particles on a line will stay on the line so long as the velocities of the particles are parallel to the line itself. This is because the magnetic force $\bold{v} \times \bold{B}$ will vanish in this situation. Hence, the collinear system is a consistent truncation of our original Lagrangian (\ref{multipLag}). This is no longer true for the coplanar case.

As was already discussed, we need at least three particles to find chaotic features. Three backreacting particles on a line have six degrees of freedom with a conserved energy, a situation closer to the triple pendulum. We can however take the mass of one of them to be much larger than the other two such that they behave as two probes in a fixed background. The probes are allowed to interact with each other since we do not restrict the $\theta_i$ and $\kappa_{ij}$ in any way, except $\sum_{i} \theta_i = 0$. The equations governing the two probes can be extracted from the two-probe Hamiltonian:
\begin{equation}
H_{col} = \frac{{p}_2^2}{2 m_2} + \frac{{p}_3^2}{2 m_3}+\frac{1}{2 m_2}\left(\theta_2+\frac{\kappa_{21}}{2 x_{21}}+\frac{\kappa_{23}}{2 x_{23}}\right)^2+\frac{1}{2 m_3}\left(\theta_3+\frac{\kappa_{31}}{2{x_{31}}}+\frac{\kappa_{32}}{2 x_{32}}\right)^2.
\label{two_lagrangian}
\end{equation}
The above Hamiltonian is a good approximation for the three-particle system in the limit where $m_1 \gg m_2, m_3$. In this limit one particle becomes non-dynamical and the energy is fully contained in the motion of the two light particles. Thus there is a conserved quantity associated to the motion of the light particles and we are left with a three-dimensional phase space, which is also the dimensionality of the double pendulum phase space.


The ground state $(x_{21}^*, x_{31}^*)$ is found by setting $U_{2} = U_{3} = 0$. In addition to imposing the triangle inequality to fully specify the ground state, we must also declare the ordering of the three particles on the line. A given ground state is mapped to a family of ground states via the scaling relation $(x_{21}^*, x_{31}^*; \kappa_{ij}) \to \lambda (x_{21}^*, x_{31}^*; \kappa_{ij})$. Slightly increasing the energy leads to small oscillations about the equilibrium position $(x_{21}^*, x_{31}^*)$. The linearized normal frequencies are the eigenvalues of the $\Omega^2$ matrix:
\begin{equation}
\Omega^2_{jl} = M^{-1}_{jk} \partial_k \partial_l H_{col}~, \quad M^{-1}_{ij} \equiv \frac{1}{m_i} \delta_{ij}~, \quad i,j,k = \{2,3 \}~.
\end{equation}
The derivatives of $H_{col}$ are evaluated at the equilibrium point. Though the general formulae for the normal frequencies are quite involved they are readily computed. As an example, the normal frequencies for $\kappa_{13} =  \kappa_{12} = - \kappa_{23} = 1$, $\theta_1 = - \theta_2 = - 1$, $\theta_3 = 0$, and $m_2 = m_3 = 1$ are:
\begin{equation}
\omega^2_\pm = \frac{2}{81}{\left( 121 \pm 13 \sqrt{73} \right)} \approx \left\{ \begin{array}{l} 5.73 \\ 0.25\end{array} \right.~.
\end{equation} 

\subsection{Poincar\'{e} Sections}

When the system is integrable or quasi-integrable, i.e. for sufficiently low energies, the trajectories in the four-dimensional phase space will reside on two-dimensional tori, since the system is simply given by two linearly coupled oscillators. Since it is hard to visualize motion on the torus, we study instead particular snapshots of the system, known as Poincar\'{e} sections (see \cite{hand,arnold} for a more complete discussion). For a given energy, we can record (over many different initial conditions) the coordinate and conjugate momentum of one particle every time the other particle has positive momentum and crosses a particular point. The resulting contours in phase space, collectively known as a Poincar\'{e} section, display the transition from quasi-integrable to chaotic behavior in our system. 

For sufficiently low energies, the Poincar\'{e} sections are given by two fixed points surrounded by a set of concentric contours. The fixed points correspond to motion in one of the two normal modes. It is useful to define the winding number $w$, which is the ratio of the number periods one particle completes for every full period completed by the other. At the linearized level away from the fixed point, the winding number $w = \omega_1 / \omega_2$. If $w$ is {\it not} a rational number, the trajectory will never quite return to its original position and thus fills one of the concentric contours. As the energy is increased, the winding number is detuned and eventually may even become rational. Hence, parts of the phase space acquire new fixed points with their own concentric contours. These correspond to nonlinear resonances. The last tori to break are those with the `most' irrational $w$ (the golden mean $(\sqrt{5}-1)/2$ is the `most' irrational number, as defined by the speed of convergence of its continued fraction expansion). The breaking of the original two islands into an increasing number is qualitatively similar to the case of a double pendulum. Eventually, there is essentially no visible structure left in the Poincar\'{e} section and we are in a regime of global chaos. 
\begin{figure}
\begin{center}
{\includegraphics[height=45mm]{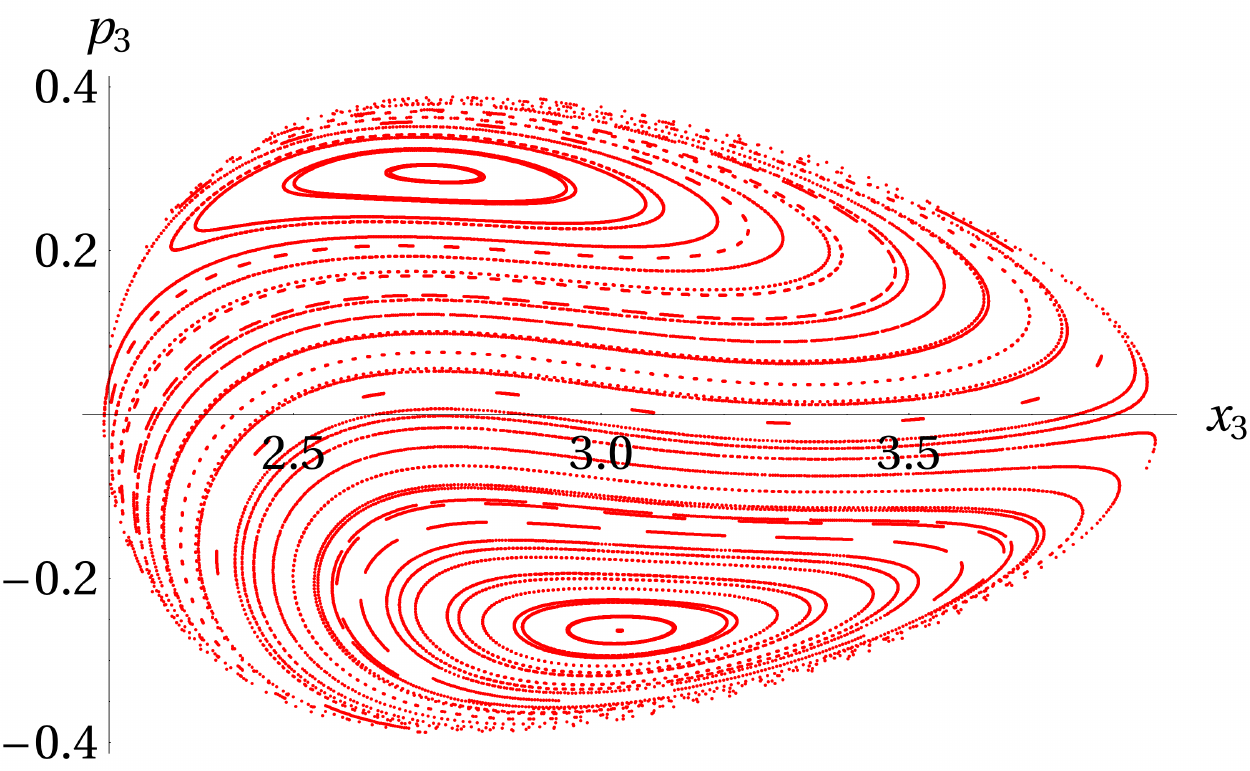} \includegraphics[height=45mm]{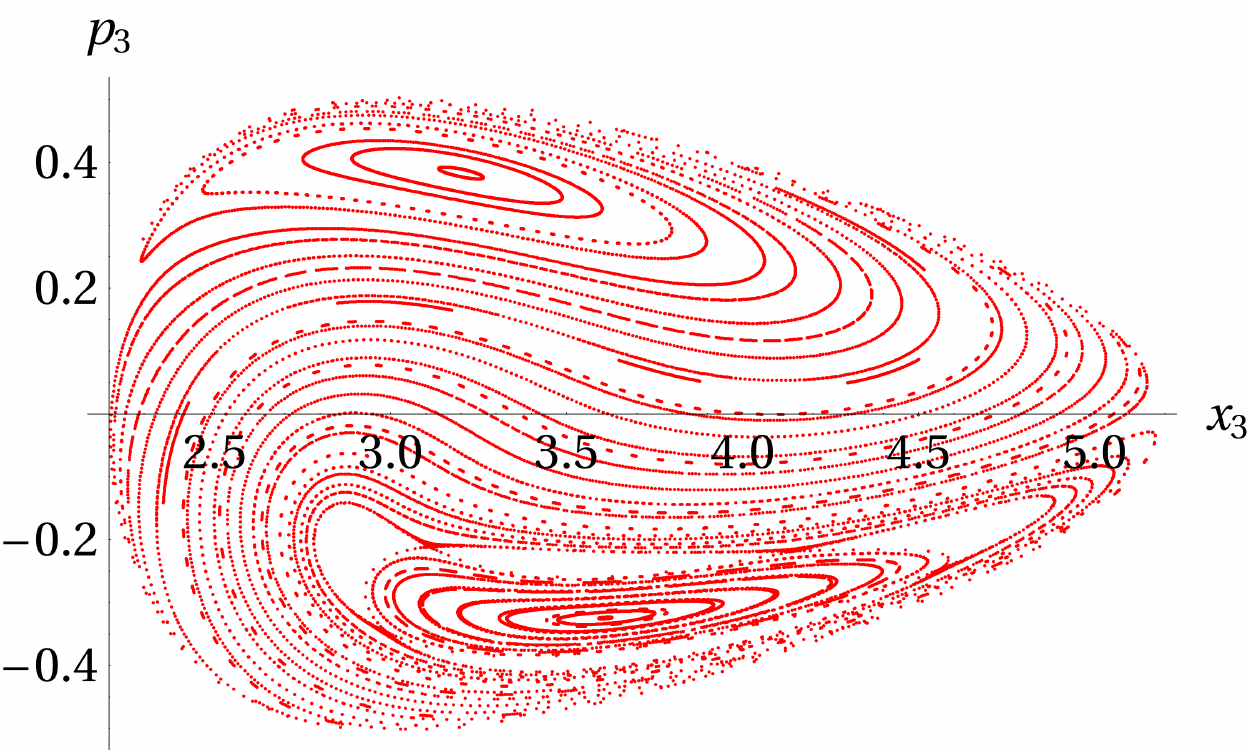}  \\
\includegraphics[height=45mm]{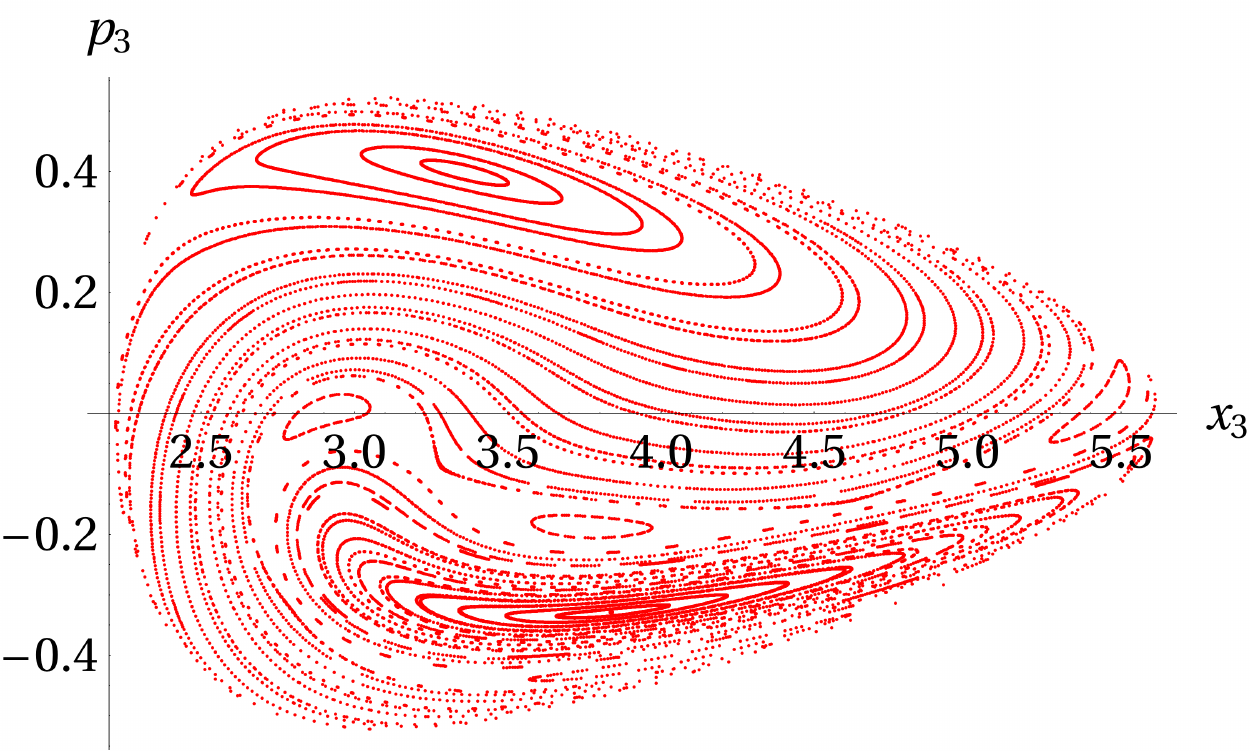} \includegraphics[height=45mm]{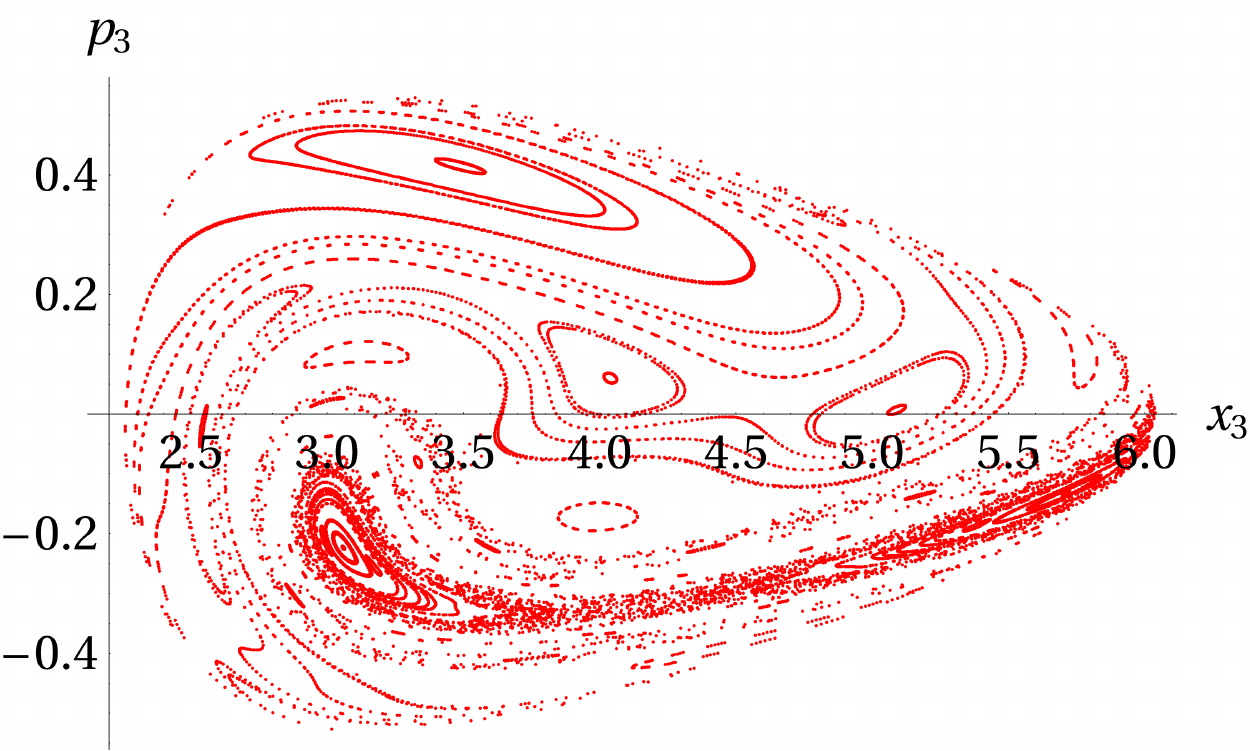} \\
\includegraphics[height=45mm]{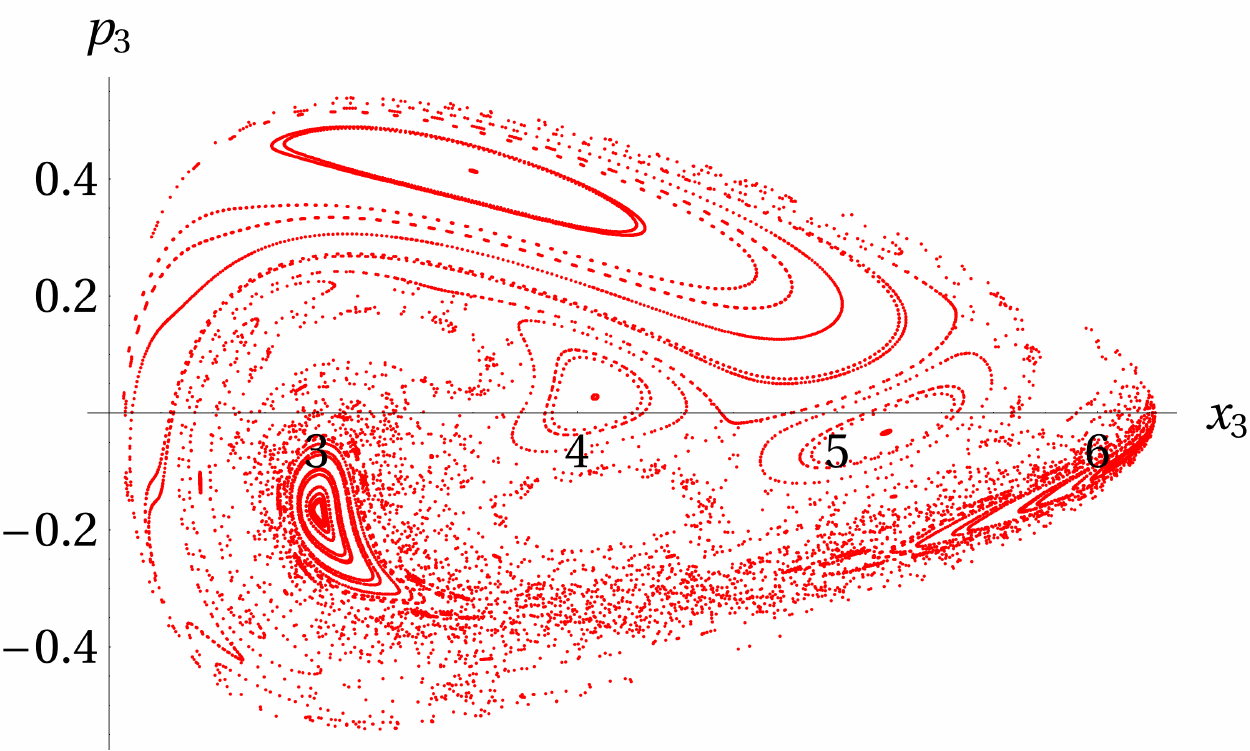} \includegraphics[height=45mm]{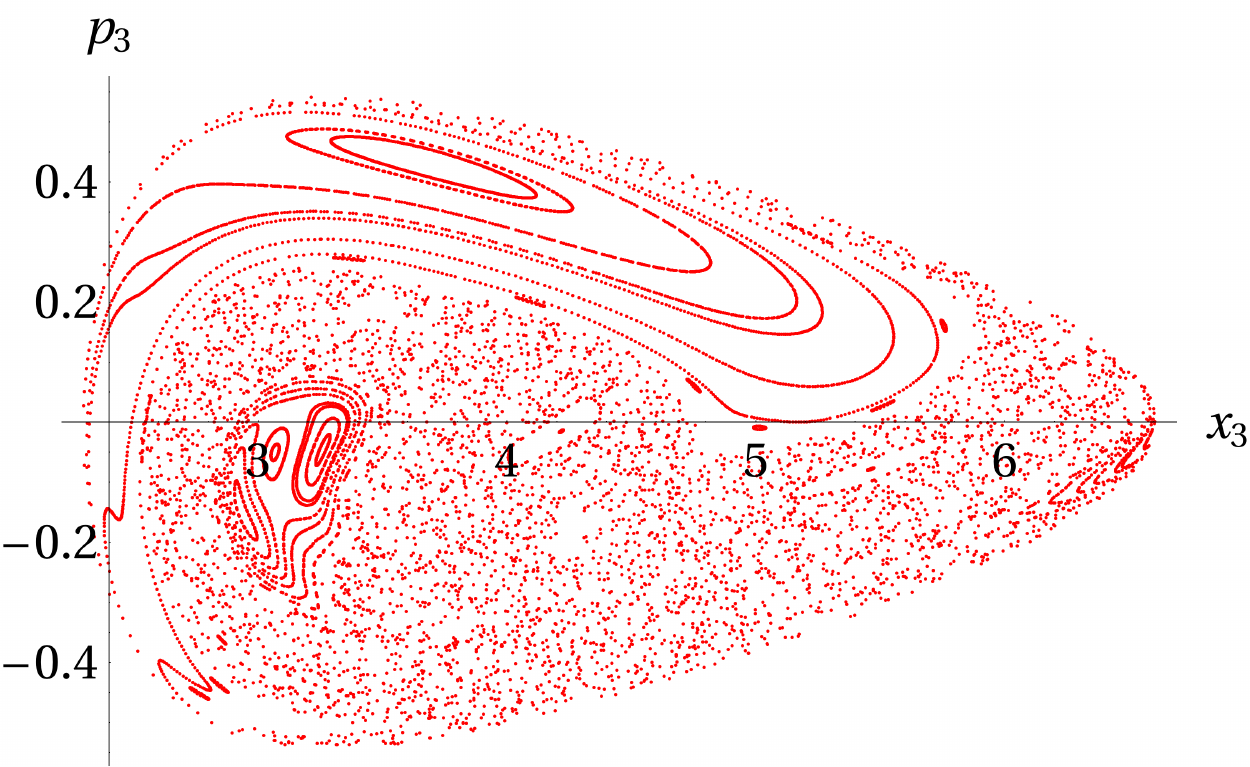}
}
\end{center}
\caption{Poincar\'{e} sections of collinear setup with $\kappa_{31} = 10$, $\kappa_{32} = -10$, $\kappa_{21} = -10$ and $\theta_3 = -1$, $\theta_2 = 1$ and energies $E = \{ 0.10, 0.20, 0.23, 0.26, 0.27, 0.30\}$. Note that the $\kappa$'s form a closed loop. The horizontal axis represents the position of particle 3 while the vertical axis represents its conjugate momentum. Any given plot is produced by varying the initial positions and momenta of the two probes subject to a fixed total energy. The pair $(x_3(t), p_3(t))$ is plotted every time the resulting trajectory of particle 2 crosses some fiducial point ($x_2(t)=x_c$) with positive momentum ($p_2(t)>0$), i.e. roughly every time particle 2 completes a full cycle as it oscillates back and forth. In the quasi-integrable regime, different initial conditions correspond to different contours. The first Poincar\'{e} section shows a quasi-integrable behavior with two fixed points corresponding to the two low energy normal modes.}\label{fig:poincaresections}
\end{figure}
We give an example of this in figure \ref{fig:poincaresections}. Similar transitions to chaos are found for examples where the $\kappa$'s form closed and non-closed loops. In several examples where the $\kappa$'s form a closed loop and obey the triangle inequality, and the $\theta$'s have the {\it same} sign, the formation of islands around fixed points representing nonlinear resonances seems to be far less manifest in the Poincar\'{e} sections. In other words, global chaos seems to set in much more quickly. We hope to study these issues systematically in the future.


\section{Trapping}

In this section, we envision a trapping problem. The setup consists of a localized bound state and another particle, which we take to be a probe, beginning inside the molecule. The probe begins its life at a random position well within the molecule. We explore the dynamical evolution of the probe as we vary the initial energy.

\subsection{Setup and Energetics}

Our setup will consist of a probe with charge $\gamma_p = (1,0)$ in the presence of a bulk molecule comprised of a number $N_c$ of fixed electric centers of charge $\gamma_c = (0,\kappa)$. The positions of the electric centers will be obtained by drawing random points from a ball of radius $R_{mol}$ using the algorithm in \cite{barthe,weisstein2}. The classical probe Hamiltonian in this background is given by:
\begin{equation}
H_{probe} = \frac{\left( \bold{p}_p - \bold{A}_p \right)^2}{2 m_p} + \frac{1}{2 m_p} \left( \sum_{i=1}^{N_c} \frac{\kappa_{p i}}{r_{pi}}  + \theta_p \right)^2~.
\end{equation}
Since all the background centers have the same charge, the $\kappa_{pi} \equiv \kappa$ are all equal. Also, to ensure that trapping occurs we require that $\kappa$ and $\theta_p$ have opposite signs. The zero energy configurations are given by setting the second term in $H_{probe}$ to zero. As usual, there is a classical moduli space $\mathcal{M}$ due to the fact that we have three probe coordinate degrees of freedom and we are solving only one equation. We could search for non-zero static minima of $H_{probe}$, but a simple computation of the gradient of the potential shows that there are only zero energy minima. The minimal energy required for the probe to reach infinity is $E_{min} = \theta_p^2/2m_p$. 

For probe energies $E_{in} \ge \theta_p^2/2m_p$ the probe can easily escape the molecule. For energies in the range $E_{in} < \theta_p^2/2m_p$, we observe trapping. Our goal is to begin quantifying the amount of classical trapping. We do this by studying the fractional volume $f_V(E_{in},t)$ covered by the probe as a function of initial energy $E_{in}$ and total trajectory time $t$. We estimate $f_V$ by studying how many centers the probe trajectory approaches to within one-half of the average inter-particle distance $r_{i.p.} \sim R_{mol}/N_c^{1/3}$. We again stress that we keep the molecule and initial positions of our probe fixed through all the trials, only varying the initial velocity of the probe. We take $m_p=1$, $\theta_p=-10$, $R_{mol}=20$, $N_c=100$ and $\kappa=1$ for the presented data.

\subsection{A trap}

\begin{figure}
\begin{center}
{\includegraphics[height=39mm]{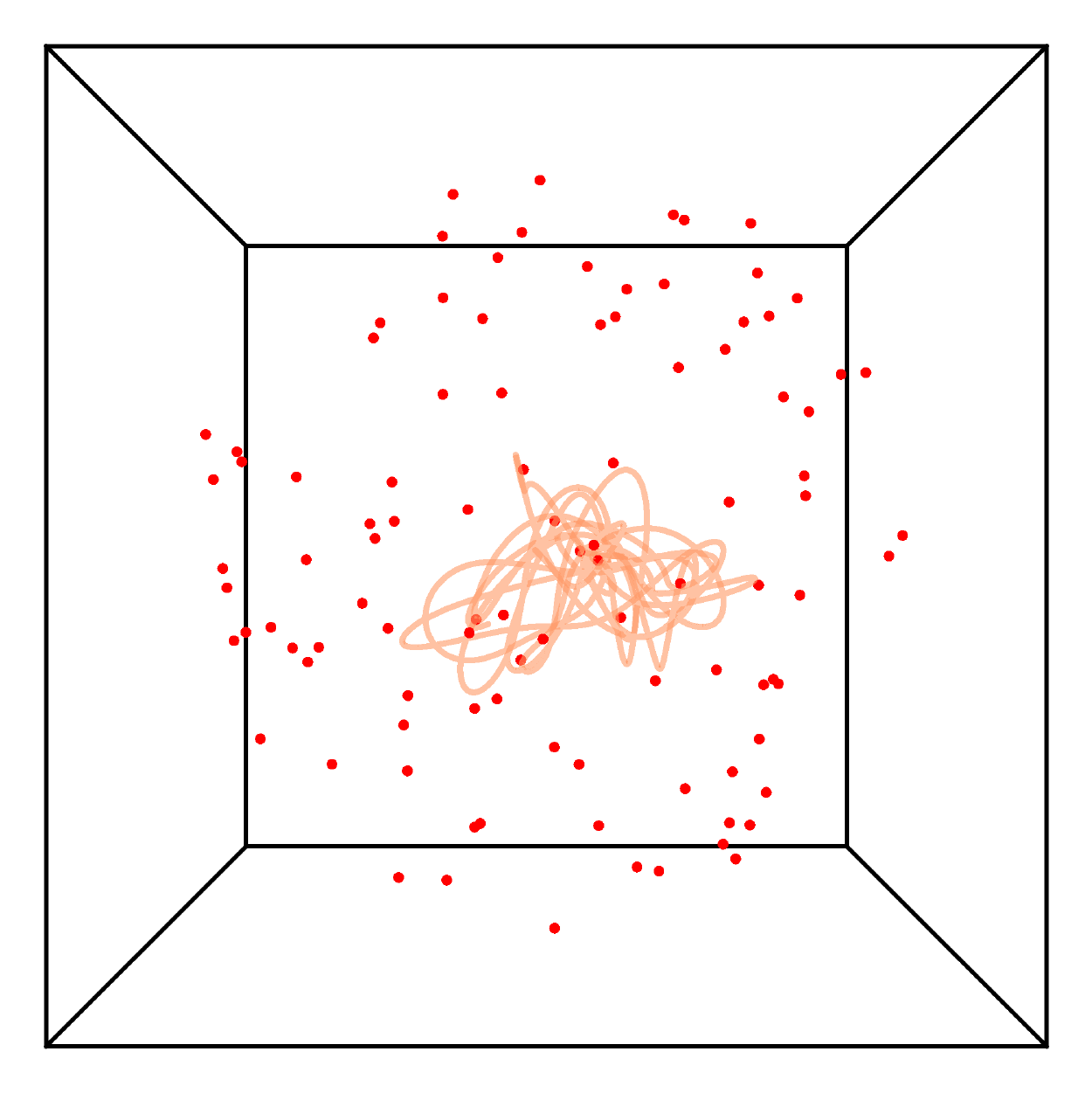}
\includegraphics[height=39mm]{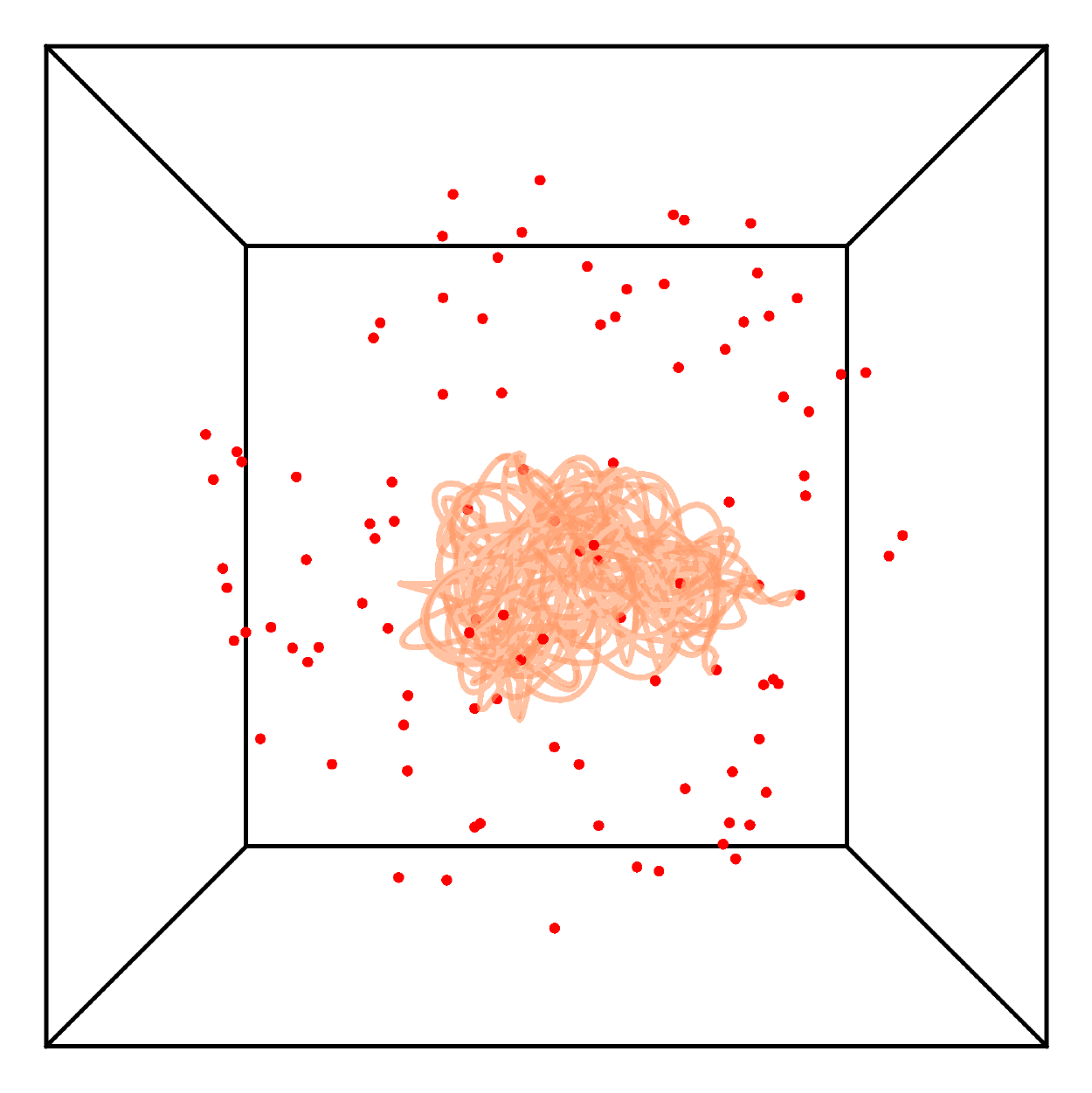}\hspace{10mm}
\includegraphics[height=38mm]{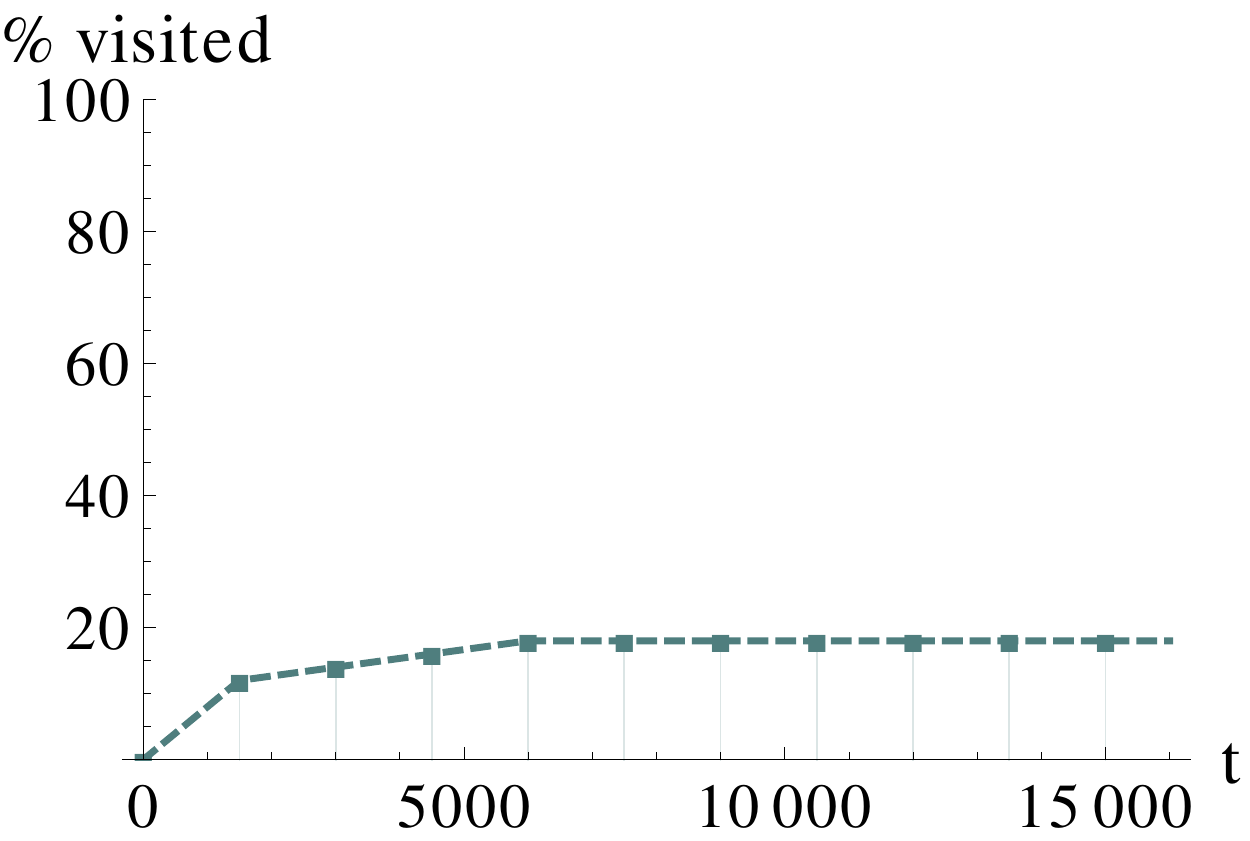}\\
\includegraphics[height=39mm]{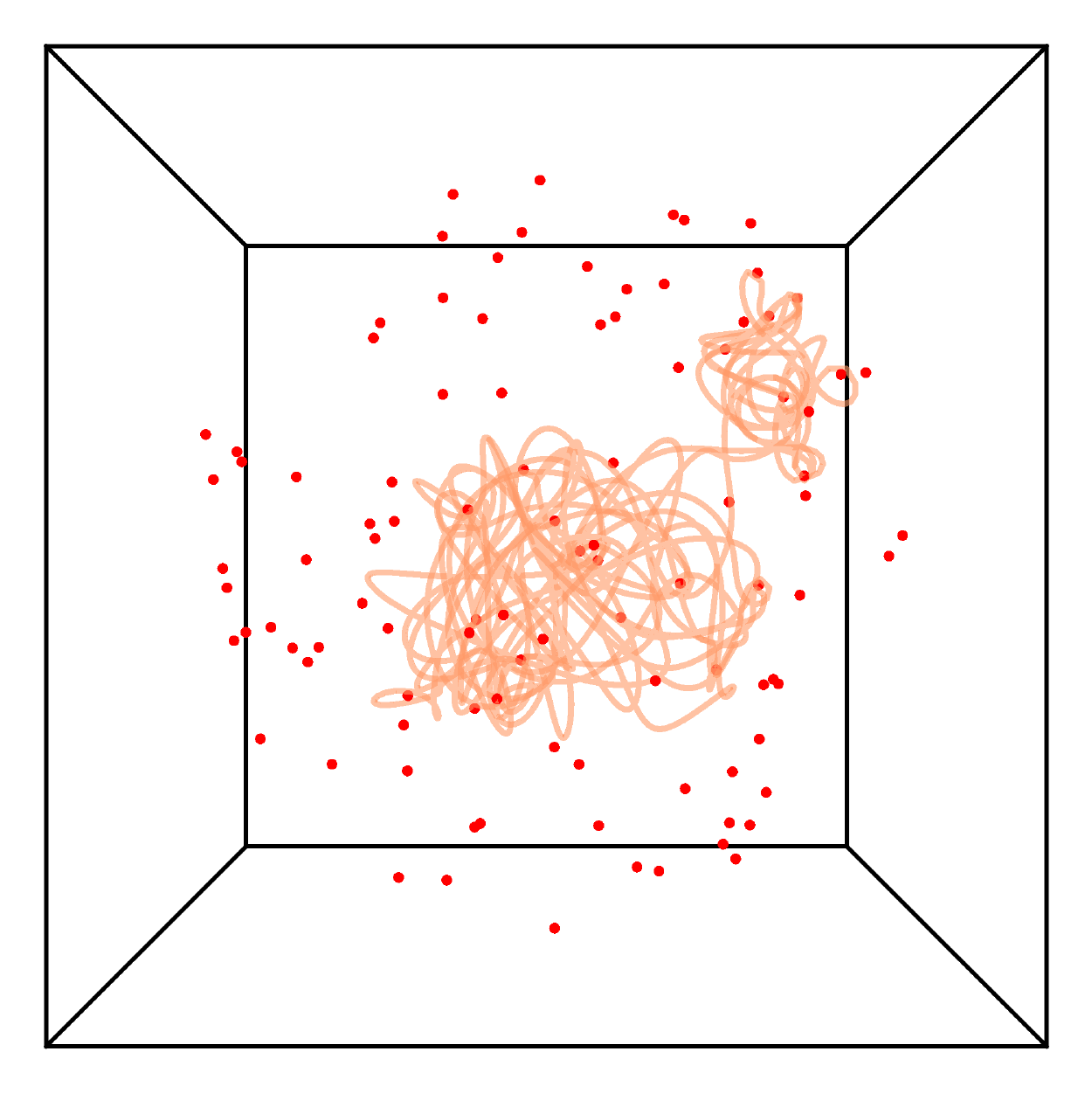}
\includegraphics[height=39mm]{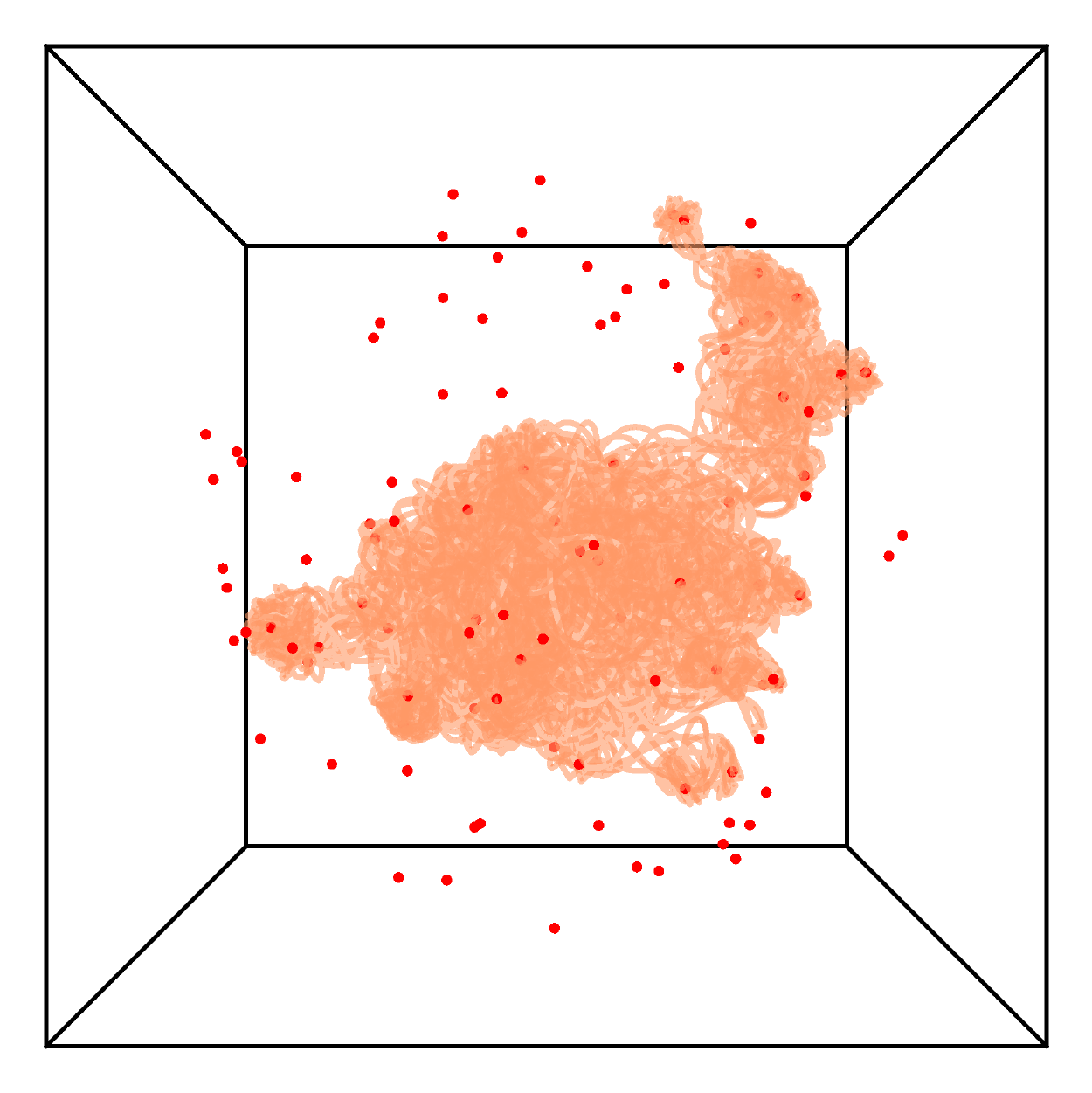}\hspace{10mm}
\includegraphics[height=38mm]{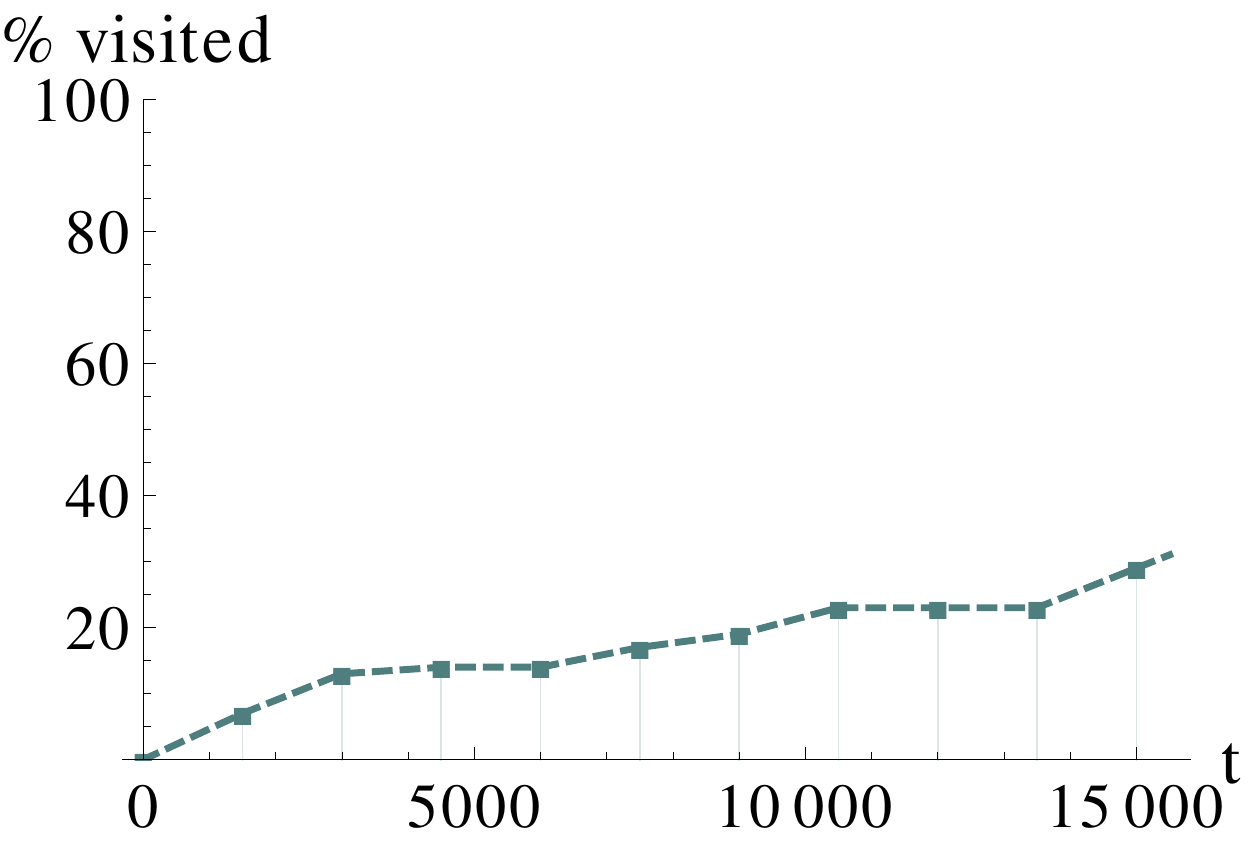}\\
\includegraphics[height=39mm]{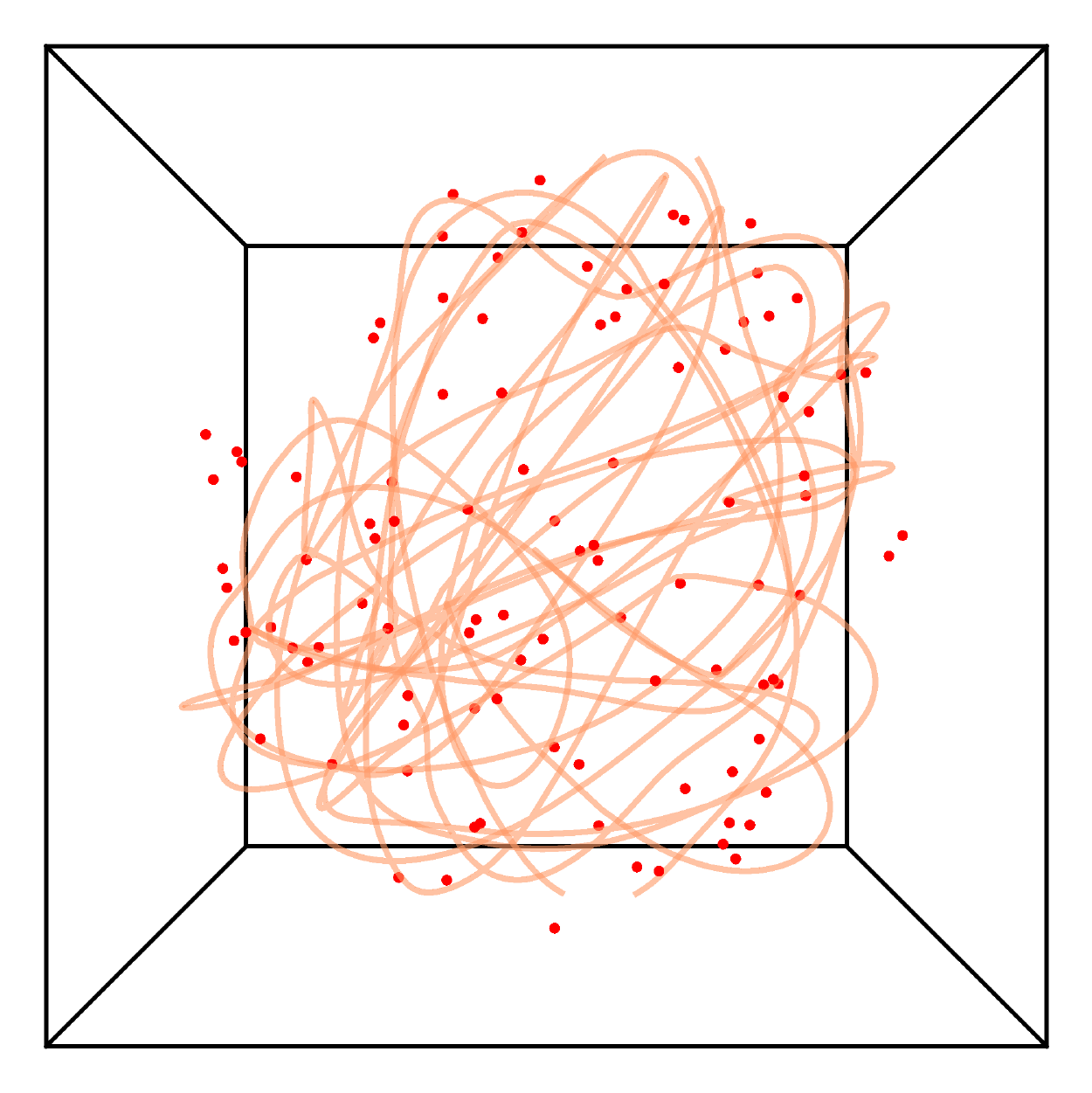}
\includegraphics[height=39mm]{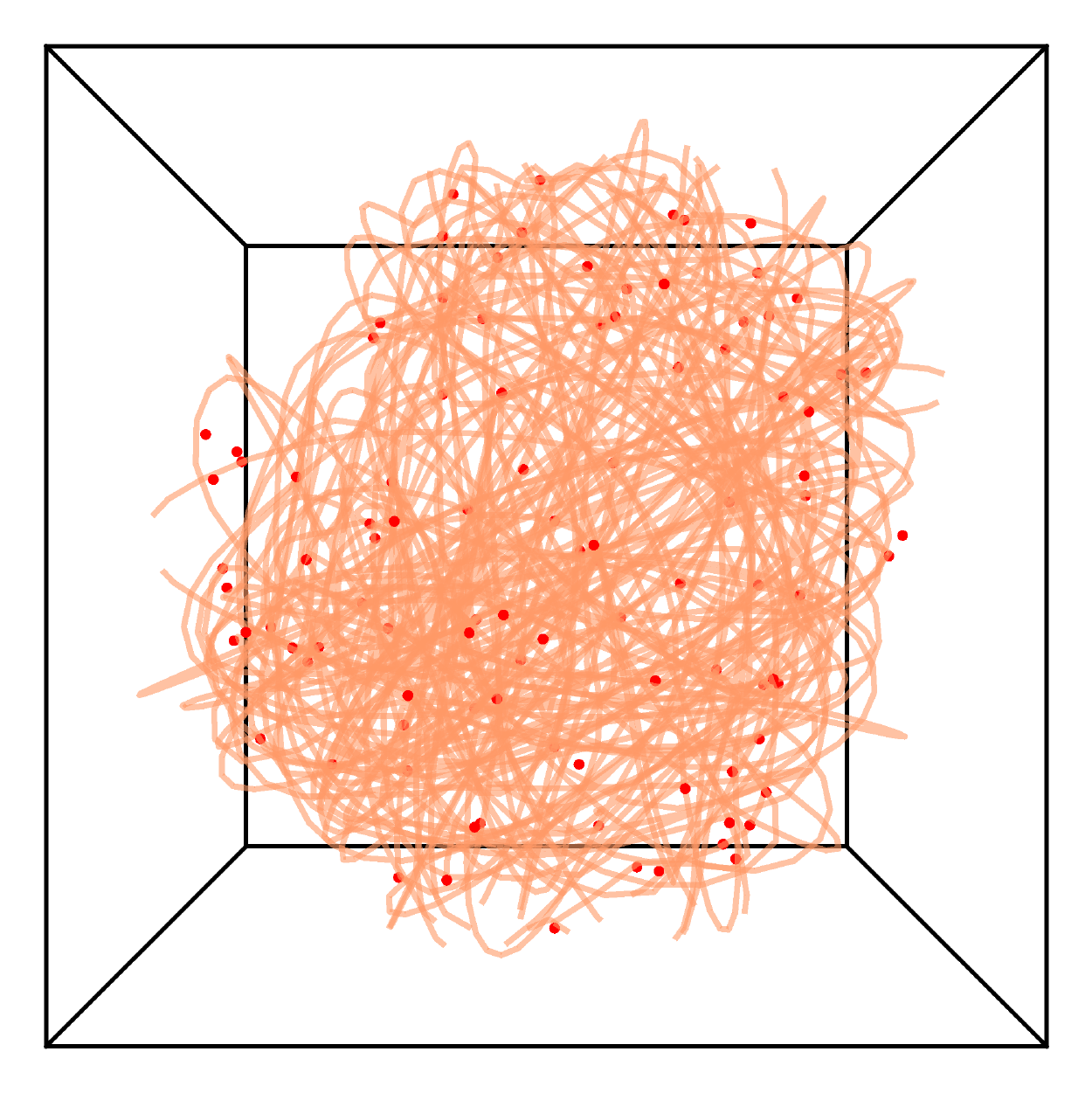}\hspace{10mm}
\includegraphics[height=38mm]{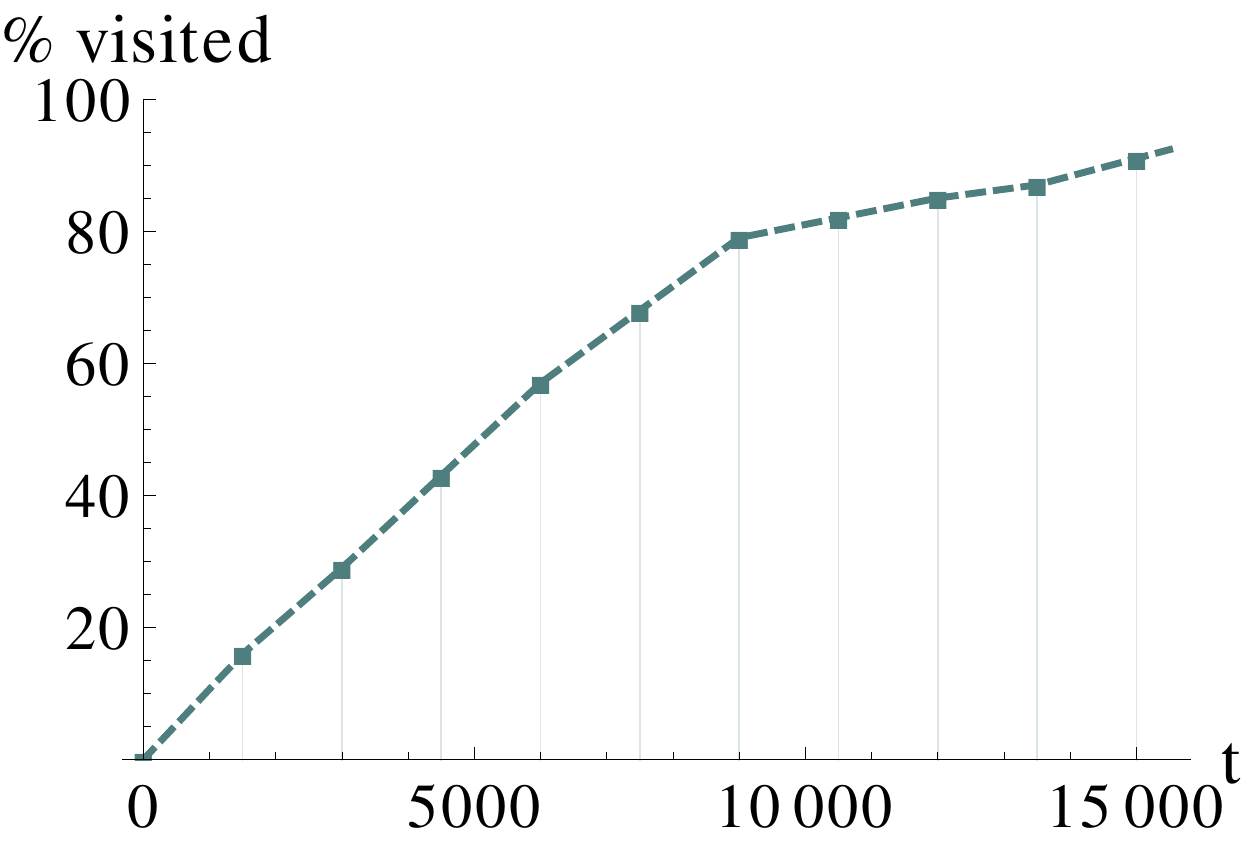}
}
\end{center}
\caption{The first row represents a low energy probe, which remains stuck in a subset of the phase space for seemingly arbitrarily long times. The second row represents an intermediate energy probe which illustrates the non-uniform escapes that occur from the low energy trapping behavior. We see that for a while it remains trapped in some subset of phase space, after which it escapes and gets stuck in some other subset of phase space. The final row represents a high energy probe which uniformly explores the molecule. The associated plots represent the percentage of the molecule explored as a function of the integration time, up to 15000 time steps in increments of 1500. The tapering off of the high energy probe is simply due to saturating the entire molecule. Below these points the increase is very uniform. The initial energy increases from 50\% of the escape energy in the first row to 60\% of the escape energy in the third row. These percentages, however, are very dependent on the parameters (e.g. $\kappa$, $\theta$, etc.) in the problem.}\label{trajectories}
\end{figure}

At low energies, we witness characteristic trapping: figure \ref{trajectories} shows one such example where a probe is confined to less than 20$\%$ of the molecule, exploring the same part of the molecule over and over again. We remind the reader that it is possible to be trapped in one region indefinitely and this behavior should not (necessarily) be looked at as a failure of not integrating for a long enough period of time. Indeed, as is seen in the Euler-Jacobi flower of figure \ref{closed}, probes can remain in one part of a molecule for arbitrarily long periods of time.  As we increase the energy, we see a transition that opens up more of phase space to the probe. Little pockets in the potential landscape form through which the probe particle can escape and begin exploring other regions of the molecule. Often this happens by sudden jumps, as illustrated in the middle row of figure \ref{trajectories}. We have tracked the energies of the probe and the numerics are stable. The jump is not due to an erroneous kick in the integrator but rather appears to be due to small pockets through which the probe can escape given enough time. Finally, at high energies (e.g. around one half of the escape energy), the probe uniformly explores the entire molecule, as illustrated in the last row of figure \ref{trajectories}.

It is also interesting to study the Schr\"{o}dinger equation for the probe in this background to see if the wavefunction exhibits trapping via some avatar of Anderson localization.\footnote{Thanks to Douglas Stanford for discussions on the quantum dynamics.}

\subsection{Topology of the potential landscape}

\begin{figure}
\begin{center}
{\includegraphics[height=50mm]{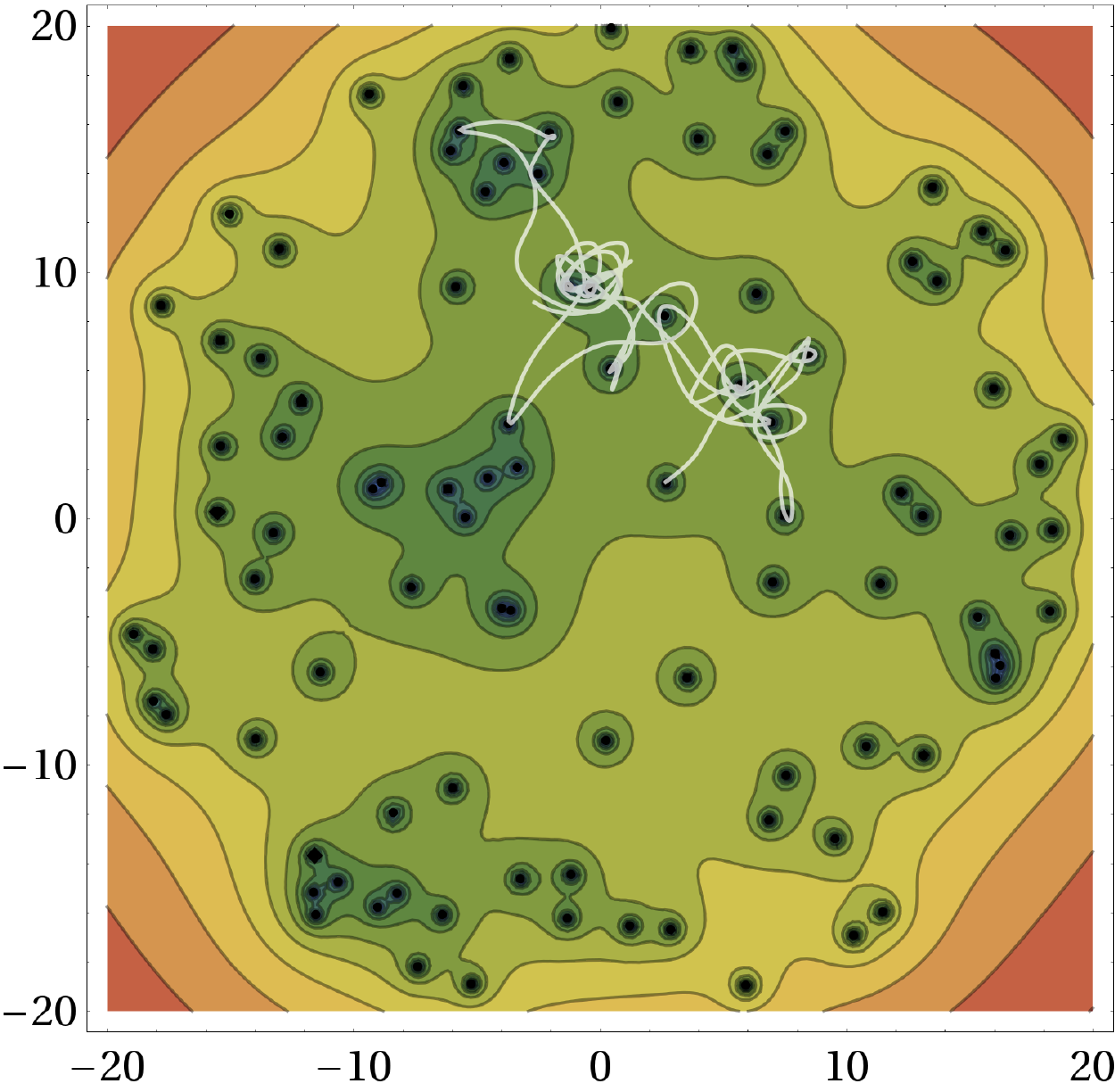}
\includegraphics[height=50mm]{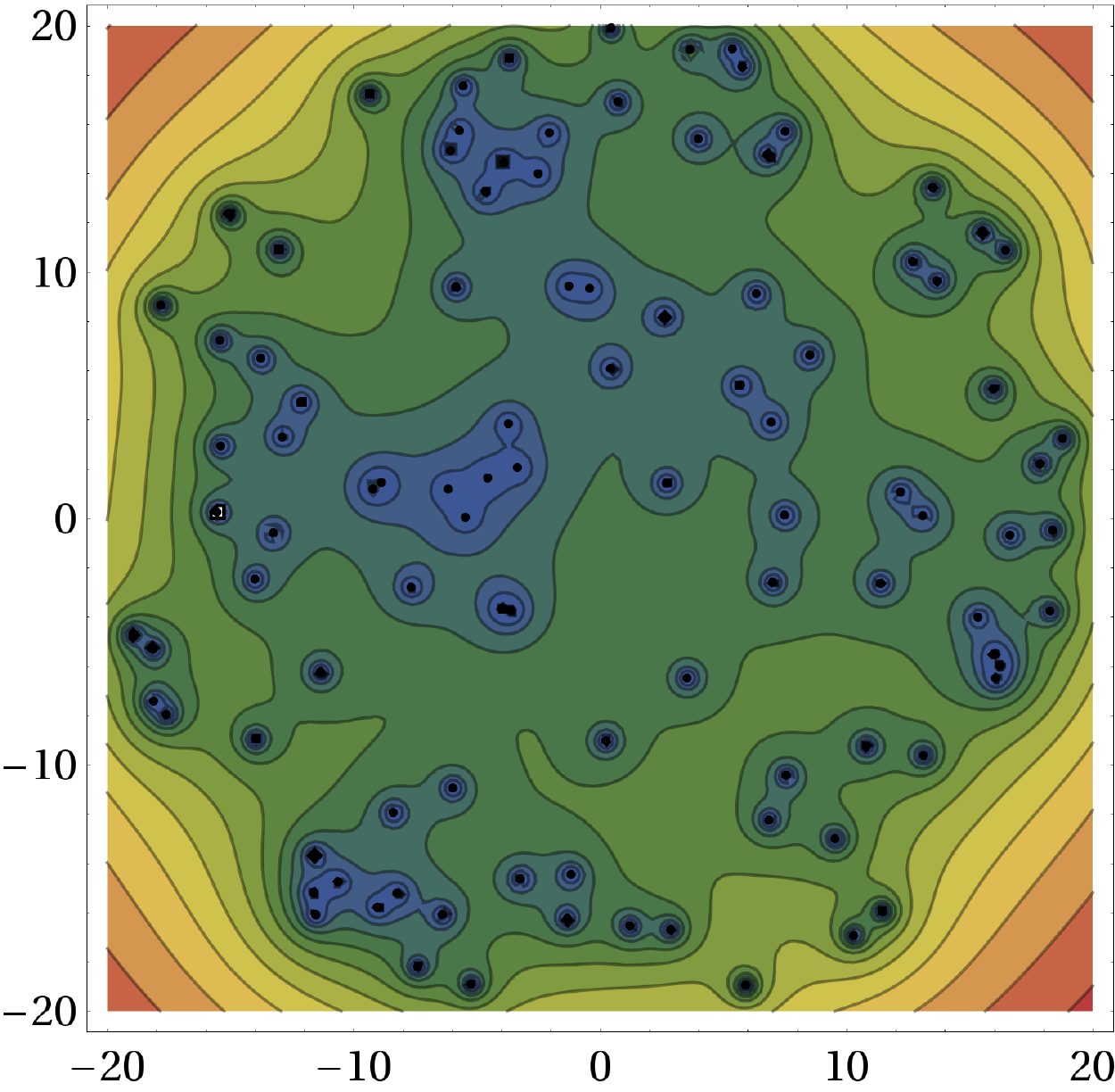}
\includegraphics[height=50mm]{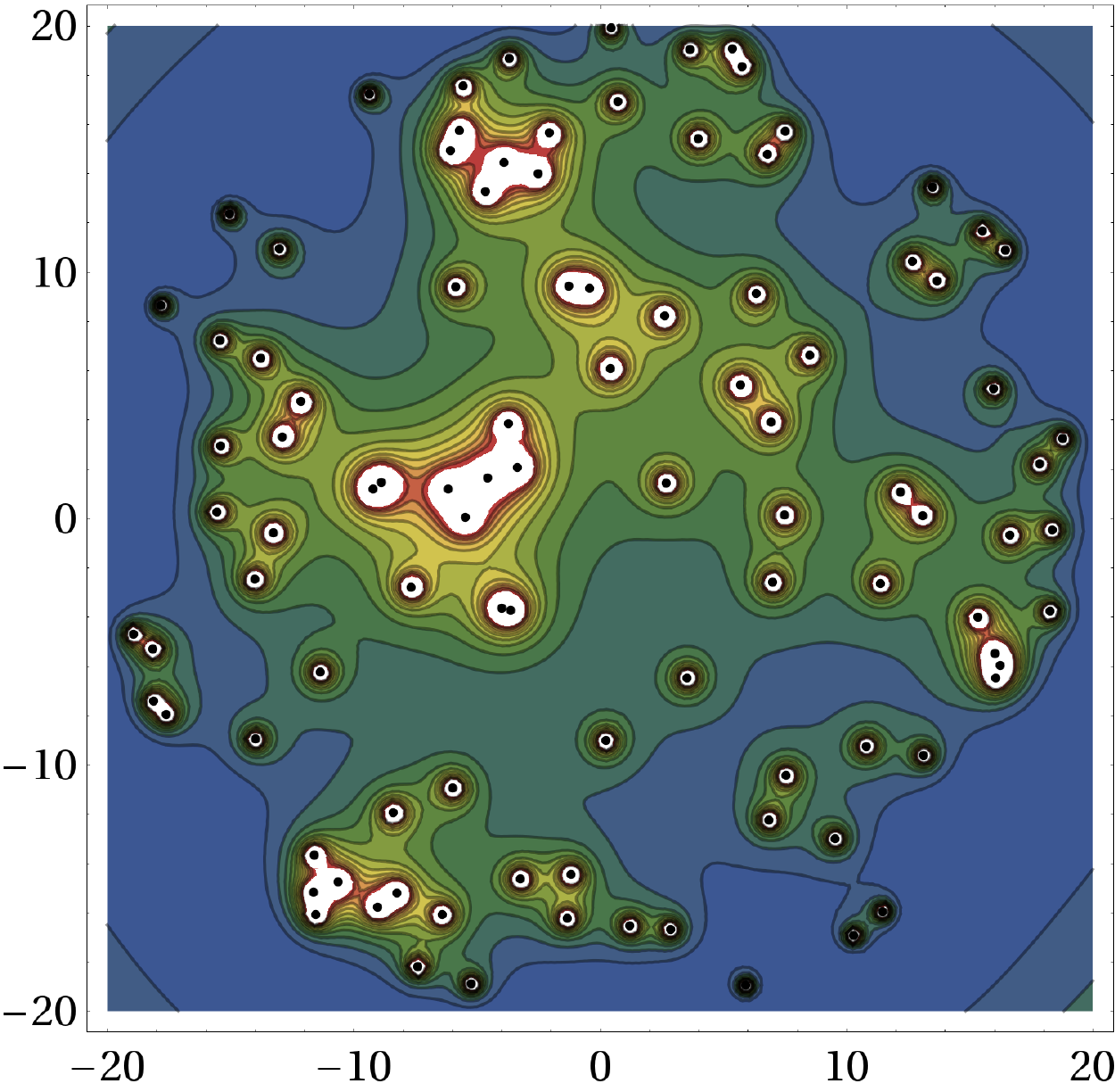}
}
\end{center}
\caption{These contour plots show equipotential surfaces in the plane of a 2D molecule consisting of one hundred centers. From left to right, we have chosen $\kappa=1$, $\kappa=1.5$, $\kappa=3.5$, and in all cases $\theta=-10$. We observe that as the magnitude of $\kappa$ increases, the minima (blue region), which initially lied near each center, are collectively expelled, forming an overall minimum that surrounds the molecule as a whole. For $\kappa=1$, the trajectory remains close to the plane of the molecule and has been superimposed on the left contour plot (transparent white line). The axes label the $x$ and $y$ positions of the probe particle.}\label{contour_plots}
\end{figure}

To illustrate the potential landscape and gain some intuition for the motion of trajectories, we set up a co-planar molecule. Again, all centers in the molecule have equal charge and their positions are chosen by uniformly selecting $N_{c}$ points on a disk of size $R_{mol}$ \cite{diskpoint}.  A probe in this background will not remain in the plane due to the magnetic fields which will push it out. However, for a molecule where every center attracts the probe, at low energies the deviations from the plane are small relative to the size of the molecule, which can be made arbitrarily large. Thus, plotting equipotential contours over this two-dimensional molecule gives an accurate picture of the potential landscape which can be used to understand the trajectories. See figure \ref{contour_plots} for such a comparison. For a fixed molecule size, as the magnitude of $\kappa$ is increased relative to the magnitude of $\theta$, the topology of the equipotentials changes by expelling the low energy part of the landscape to the outside of the molecule, as can be seen in figure \ref{contour_plots}. This mimics the change in topology of the moduli space in going from Region V to Region IV in figures \ref{phaseplot} and \ref{regions}. Topology changes in the moduli space of the three center setup also occurs as $\kappa$ is increased while keeping $\theta$ fixed as well as the distance between the background centers.

\section{Holography of Chaotic Trajectories?}

We end our journey by discussing how the picture we are developing may fit into the broader context of holography. The usual interpretation of a large black hole in an asymptotically anti-de Sitter space is that we have prepared the dual CFT in some finite temperature state. On the other hand, the presence of a vast number of distinct entropically relevant multicentered black hole configurations inside an anti-de Sitter universe \cite{adshothalos} implies a vast number of minima in the free energy of the dual CFT (as a function of configuration space). In particular, the usual assumptions of the no-hair theorem fail since a set of macroscopic charges does not uniquely fix the bulk solutions. In fact, the solutions are characterized by a large collection of multipole moments. Furthermore, though most of them do not constitute true ground states, they can be very long lived, decaying mostly through thermal and quantum tunneling. 

We can access information about the relaxation and response of the CFT by computing boundary-to-boundary correlators in the bulk. In the large frequency (or large mass) limit two-point functions have been associated to bulk geodesics which begin and end their life near the boundary of AdS \cite{Louko:2000tp,Fidkowski:2003nf,Festuccia:2005pi,Kraus:2002iv}. Such geodesics will become highly complex and chaotic in the bulk due to the presence of the non-trivial black molecule, as evidenced by our simpler setup. In fact, a geodesic may become trapped in some very long lived unstable orbit before escaping back to the boundary. Thus, the  two-point function expressed as a path integral over bulk trajectories and the applicability of the saddle point approximation may be a somewhat involved issue. 
This picture suggests that the {\it linear response} properties of the dual CFT, to the extent that they are captured by the two-point function in the geodesic approximation, in the multicentered/glassy phase are rather different from those in a usual thermal state, where for example the motion of geodesics is integrable.  The motion of a very massive probe or high energy graviton falling into the bulk corresponds to a point-like source cascading to lower energies (and covering larger size) in the CFT. Eventually the excitation returns back to a point-like source at some other point on the two-sphere where the CFT resides. From the bulk point of view this is when the particle dropped into the molecule returns back out. The chaotic nature of the bulk physics suggests chaotic behavior of the boundary theory itself.


One may also consider the dynamical features at zero temperature for which asymptotically AdS$_3$ multicentered configurations are known \cite{deBoer:2008fk}. The possible presence of {\it classical} chaotic behavior of the bulk AdS$_3$ should correspond to {\it quantum} dynamics in the dual CFT$_2$. One effect of particular interest in chaotic systems is known as {\it quantum scarring}, where it has been observed that the wavefunction of a chaotic system peaks on \emph{closed} classical trajectories \cite{heller}. We hope to explore these issues further in future work.

\section*{Acknowledgements}

It is a great pleasure to acknowledge Sean Hartnoll, Finn Larsen, Leo Pando-Zayas, Lucas Peeters and Steve Shenker for useful discussions. This work has been partially funded by DOE grant DE-FG02-91ER40654 and by a grant of the John Templeton Foundation. The opinions expressed in this publication are those of the authors and do not necessarily reflect the views of the John Templeton Foundation. T.A. would like to thank the Institute for Theoretical Physics at KU Leuven, the Stanford Institute for Theoretical Physics, and the Michigan Center for Theoretical Physics for their kind hospitality during the completion of this work. 

%
%

\appendix


\section{Two-Body Problem}\label{sec:twoparticles}

In this appendix we discuss the motion of a single probe particle in the background of a fixed charge sitting at the origin. This was studied at length in \cite{Avery:2007xf}. The Hamiltonian of this system is given by 
\begin{equation}\label{twopartham}
H=\frac{1}{2 m}(\mathbf{p}-\mathbf{A})^2 +\frac{1}{2m}\left(\frac{\kappa}{2r}+\theta\right)^2~,\quad \mathbf{p}\equiv m\dot{\mathbf{x}}+\mathbf{A}
\end{equation}
and is conserved. For simplicity we choose $\kappa>0$ and allow $\theta$ to be either positive or negative. Other than the Hamiltonian, this system admits two vector-valued conserved quantities known as the angular momentum $\mathbf{L}$ and the Runge-Lenz vector $\mathbf{n}$. Explicitly
\begin{equation}\label{eq:twopartconservedquant}
\mathbf{L}=\mathbf{x}\times(\mathbf{p}-\mathbf{A})+\frac{\kappa}{2r}\mathbf{x}~,\quad\text{and}\quad\mathbf{n}=\left(\mathbf{x}+\frac{1}{\theta}\,\mathbf{L}\right)\times(\mathbf{p}-\mathbf{A})~.
\end{equation}
Since this system is superintegrable, the probe particle's trajectories can be found algebraically. 

First, notice that $\mathbf{n}\cdot\dot{\mathbf{x}}=0$ implying that $\mathbf{n}$ is perpendicular to the plane of motion of the probe particle. We use this fact to orient our axes such that $\mathbf{n}=|\mathbf{n}|\hat{z}$. It is straightforward to show that
\begin{equation}
|\mathbf{n}|=\sqrt{\frac{2mH}{\theta^2}\left(\mathbf{L}^2-\frac{\kappa^2}{4}\right)}~,
\end{equation}
which implies that $|\mathbf{L}|\geq {\kappa}/{2}$. With this choice of coordinates, the particle's trajectory is constrained to lie in a plane of constant $z$. The magnitude of $z$ can be obtained by computing $\mathbf{n}\cdot\mathbf{x}=|\mathbf{n}|z=-\left(\mathbf{L}^2-\frac{\kappa^2}{4}\right)/{\theta}$, giving 
\begin{equation}
z=-\frac{|\theta|}{\theta}\sqrt{\frac{\mathbf{L}^2-\frac{\kappa^2}{4}}{2mH}}~.
\end{equation}
We have yet to choose an orientation for the $x-y$ plane; we do so by aligning our coordinates such that $L_y=0$ and $L_x$ points in the positive x direction. The components of the angular momentum are given by
\begin{equation}
L_x=\sqrt{\mathbf{L}^2-\frac{\theta^2}{2mH}\left(\mathbf{L}^2-\frac{\kappa^2}{4}\right)}~,\quad L_z=\sqrt{\frac{\theta^2}{2mH}\left(\mathbf{L}^2-\frac{\kappa^2}{4}\right)}~.
\end{equation}

We can determine the particle's trajectory explicitly by noticing that~(\ref{eq:twopartconservedquant}) implies that $\mathbf{L}\cdot\mathbf{x}={\kappa\,r}/{2}$ or
\begin{equation}
\left(1-e^2\right)(x-x_0)^2+2\ell e(x-x_0)+y^2=\ell^2~,
\end{equation}
which is the equation for a conic section in cartesian coordinates. The quantities $e$ and $\ell$ are the eccentricity and the semi-latus rectum of the conic section respectively and are given by
\begin{equation}
e=\frac{2L_x}{\kappa}=\frac{2}{\kappa}\sqrt{\mathbf{L}^2-\frac{\theta^2}{2mH}\left(\mathbf{L}^2-\frac{\kappa^2}{4}\right)}~,\quad\ell=\frac{4\mathbf{L}^2-\kappa^2}{\kappa\sqrt{8mH}}~.
\end{equation}
The quantity $x_0 \equiv {\sqrt{\mathbf{L}^2-\frac{\theta^2}{2mH}\left(\mathbf{L}^2-\frac{\kappa^2}{4}\right)}}/({|\theta|-\sqrt{2mH}})$ is the location of one of the foci of the conic section. 

The overall shape of a conic section is determined by its eccentricity, with elliptic orbits corresponding to $e<1$, while parabolic and hyperbolic orbits correspond to $e=1$ and $e>1$ respectively. Intuition predicates that bound orbits should only happen for $\theta<0$, while $\theta>0$ gives rise to parabolic or hyperbolic orbits. For positive $\theta$ the Hamiltonian is bound such that $2mH\geq\theta^2$ which implies that $e\geq 1$, thus verifying our intuition.

For $e<1$ the length of the semi-major axis $a$ of the elliptic orbit is given by
\begin{equation}
a=\frac{2\ell}{1-e^2}\sim\frac{\kappa\sqrt{H/2m}}{H_{\rm escape}-H}+\mathcal{O}\left({1}\right)
\end{equation}
so as the energy approaches the escape energy $H_{\rm{escape}}=\theta^2/2m$ (or as $e$ approaches $1$), the size of the bound orbit diverges. 
\begin{figure}
\centering{
\includegraphics[height=60mm]{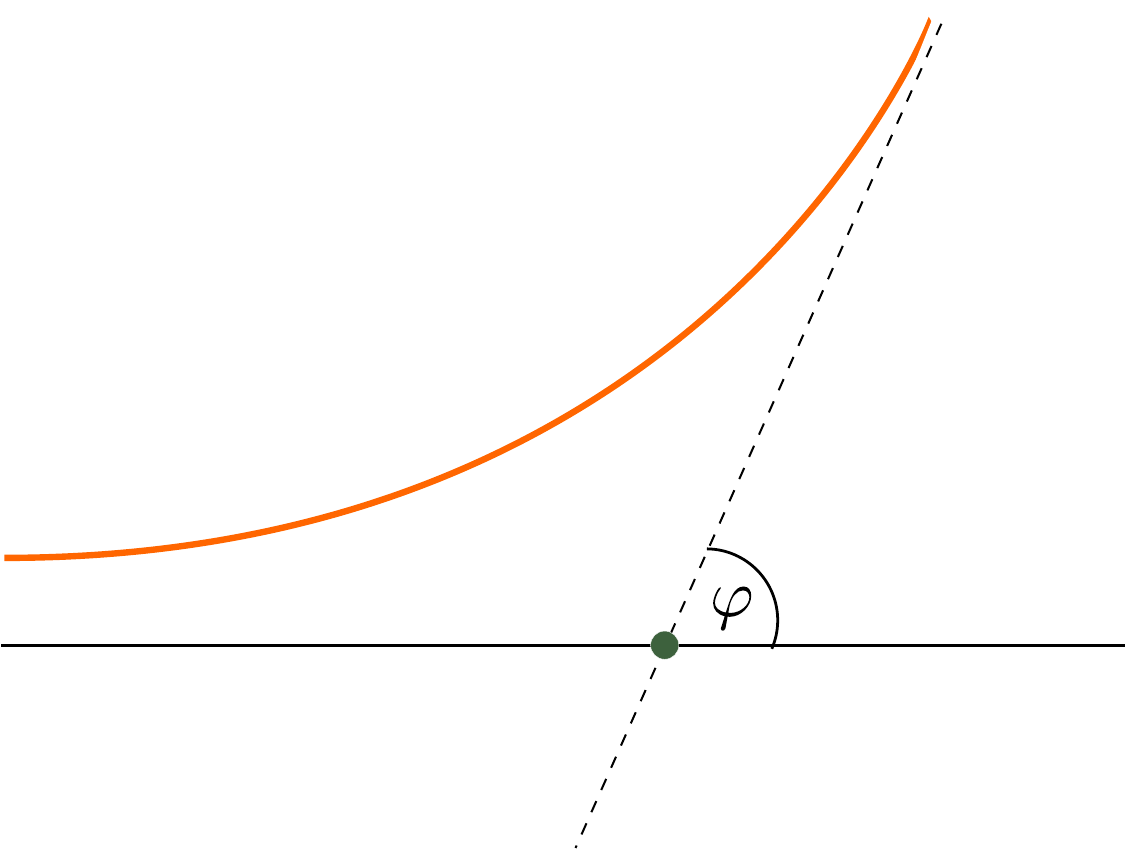}
}\caption{Scatering angle.}\label{scatter}
\end{figure}
We end this appendix with a discussion of scattering for $e\geq 1$. Since the trajectory of the particle is given by a conic section in the $x-y$ plane, the scattering angle as defined in figure~\ref{scatter} is given by
\begin{equation}
\varphi=2\,\text{arccot}\left(\sqrt{e^2-1}\right)~.
\end{equation}
For parabolic orbits ($e=1$), $\varphi=\pi$ and we see that the particle completely back scatters. As we increase $e$ the scattering angle decreases monotonically.

\end{document}